\documentclass[lettersize,journal]{IEEEtran}
\usepackage{amsmath,amsfonts}
\usepackage{algorithmic}
\usepackage{array}
\usepackage[caption=false,font=normalsize,labelfont=sf,textfont=sf]{subfig}
\usepackage{textcomp}
\usepackage{stfloats}
\usepackage{url}
\usepackage{verbatim}
\usepackage{graphicx}
\usepackage[ruled,linesnumbered,noend]{algorithm2e}
\usepackage{mdwmath}
\usepackage{mdwtab}
\usepackage{eqparbox}
\usepackage{booktabs}
\usepackage{url}
\usepackage{subfloat}
\usepackage{tabulary}
\usepackage{epstopdf}
\usepackage{multirow}
\def\BibTeX{{\rm B\kern-.05em{\sc i\kern-.025em b}\kern-.08em
    T\kern-.1667em\lower.7ex\hbox{E}\kern-.125emX}}
\usepackage{balance}
\newcommand{\system}{$\mathsf{VC4ASL}$\xspace}
\newcommand{\RN}[1]{%
	\textup{\lowercase\expandafter{\romannumeral#1}}%
}

\begin{document}
\title{Enabling American Sign Language Communication Under Low Data Rates}
\author{Panneer Selvam Santhalingam, Swann Thantsin, Ahmad Kamari, Parth Pathak, Kenneth DeHaan
\thanks{Panneer Selvam Santhalingam and Swann Thantsin are with Brooklyn College, City University of New York.\\
Parth Pathak and Ahmad Kamari are with George Mason University.\\
Kenneth Dee Hann is with Gallaudet University.\\
Corresponding author:Panneer Selvam Santhalingam}}

\markboth{Journal of \LaTeX\ Class Files,~Vol.~18, No.~9, September~2020}%
{How to Use the IEEEtran \LaTeX \ Templates}

\maketitle

\begin{abstract}
In recent years, video conferencing applications have become increasingly prevalent, relying heavily on high-speed internet connectivity. When such connectivity is lacking, users often default to audio-only communication, a mode that significantly disadvantages American Sign Language (ASL) users, whose communication relies on hand gestures, body movement, and facial expressions. In this work, we introduce VC4ASL, a system designed to enable ASL communication over the audio channel of existing video conferencing applications, even in the absence of reliable video. VC4ASL integrates seamlessly with current platforms without requiring any modifications. Our approach establishes a communication channel through audio by encoding and transmitting human pose information, which is then rendered to reconstruct signed content. We propose novel receive-side error detection and correction mechanisms that exploit the inherent structural constraints of human pose data. To evaluate the system, we simulate network-degraded environments, generate pose-based ASL video sequences, and conduct user studies to assess comprehension among ASL users. Experimental results demonstrate that VC4ASL effectively facilitates intelligible ASL communication over audio in low-bandwidth scenarios where video transmission is impaired.
\end{abstract}

\section{Introduction}
\label{sec:introduction}

We live in an era of high-speed connectivity, where average internet speeds worldwide have significantly increased \cite{internet_speed}. Access to faster networks and the growing popularity of remote collaboration have contributed to the widespread use of video conferencing and video calling applications \cite{vca_application}. However, such high-speed access is often limited to specific geographic regions and settings. Within the US, there are still areas \cite{rural} with limited broadband coverage, and mobile environments further reduce the effective data rates of cellular networks \cite{mobility}. Even where high-speed internet is available, fluctuations in network quality and the adaptive responses of video conferencing applications (VCAs) can degrade the user experience \cite{dont_forget_user,measurement_webex}. When high-speed connections are unavailable, users typically disable video and rely solely on audio. Even when video is enabled, quality degradation often goes unnoticed because audio remains continuous. Due to its lower bandwidth demands and dedicated streams \cite{webrtc}, audio often remains seamless. At the same time, video lags or freezes \cite{measurement_IMC}, and most users readily switch to audio and continue communicating.

However, such a switch to audio is not an option for deaf and hard-of-hearing individuals who communicate using American Sign Language (ASL). ASL is a visual-spatial language that relies on hand movements, body posture, and facial expressions to convey meaning \cite{ASL_2}. In the US, there are over five hundred thousand ASL speakers \cite{ASL_3}. When ASL users engage with non-ASL speakers, video relay services with interpreters are used, with interpreters facilitating communication. A persistent issue in these services is the dependency on high-quality video and fast internet connections \cite{vrs}. When sufficient data rates are unavailable, ASL users often turn to asynchronous communication methods, such as video messaging or text-based alternatives.

Existing research examining the challenges faced by ASL speakers in video conferencing contexts has primarily focused on accessibility issues such as screen size limitations, the involvement of interpreters, and the inability to interrupt, largely because most VCAs prioritize audio cues \cite{asl_accessibility_1,asl_accessibility_2,asl_accessibility_4}. On the other hand, studies evaluating VCA performance and user experience typically use talking heads \cite{measurement_IMC}, with little to no body movements, to simulate standard video conferencing scenarios, which fail to reflect the video dynamics involved in ASL. Although some early works have studied methods to support ASL communication under low data rate scenarios \cite{asl_low_data_rate1,asl_low_data_rate2}, interest in this line of research has waned. Meanwhile, advances in computer vision have produced techniques that can extract detailed human skeleton pose information, which can be used for a range of applications, from image generation to avatar animations \cite{pose_use1,pose_use_2}. We believe that these recent developments offer a promising direction for enabling American Sign Language (ASL) communication in low-data-rate environments.

In this work, we propose a system called \system, short for Video Conferencing for ASL, designed to facilitate ASL communication under low data rate conditions. Our solution is compatible with existing VCA platforms such as Zoom and Google Meet and requires no modifications to these applications. The key insight behind \system is that most ASL users, though not all, do not rely on the audio stream within video conferencing environments. As previously noted, audio streams offer a reliable communication channel under constrained bandwidth conditions, owing to their low data requirements and the availability of mature audio compression techniques. To leverage this channel, \system must operate within similarly low data rate constraints, limiting the amount of information that can be transmitted. Furthermore, audio codecs, depending on their configuration, may distort or degrade the audio signal, resulting in potential information loss. This issue has been well-documented in the audio steganography literature \cite{audio_steg_survey}.

Our approach involves encoding human pose information \cite{openpose,mediapipe} into audio, capturing the essential visual cues for ASL communication, transmitting it through the VCA’s audio stream on the sender side, and rendering the decoded information in real-time on the receiver side.We next briefly summarize the major research challenges we address and our key contributions:

\textbf{Robustness to codec compression:} Audio codecs employ compression techniques in both time and frequency domains and dynamically adjust bitrate and frame size in response to network conditions, which can corrupt encoded pose data.
We propose a novel audio frame structure based on chirp spread spectrum modulation to encode pose data into audio. In addition, we design and implement the complete signal processing pipeline required to identify frames within the audio stream, segment the signal, and recover the encoded pose data in real-time. Our approach enables reliable data transmission over audio streams of existing VCAs, yielding low symbol error rates.

 \textbf{Operating under limited bandwidth:} The low bandwidth allocated for audio necessitates highly efficient encoding of pose data to maintain an adequate poses-per-second rate and ensure a satisfactory user experience.
To address this, we adopt quantization and efficient bit-packing strategies to minimize pose size. We further introduce a temporal differencing method that transmits pose displacements and reconstructs full poses at the receiver, significantly reducing the data rate while increasing frame rate. Additionally, we exploit the structural dependencies among human joints to transmit only a subset of keypoints, and propose a lightweight model to predict the excluded joints from the received data.

 \textbf{Pose corruption due to network fluctuations:} Operating at low bitrates and under unstable network conditions increases the likelihood of pose data corruption.
To mitigate this, we propose a novel error detection mechanism at the receiver using autoencoders. We leverage the reconstruction loss between the generated and received poses as a proxy for identifying out-of-distribution samples, thereby reframing error detection as an out-of-distribution detection problem. Furthermore, we model pose corruption as a stochastic process and incorporate this into the training of our pose prediction model, improving robustness against corrupted pose inputs.

\textbf{Evaluating the performance of \system:} We conduct experimental studies to assess the \system modulation scheme by simulating varying network conditions between sender and receiver across two major video conferencing applications: Zoom and Google Meet. Our results show that the proposed modulation scheme achieves an average symbol error rate of 11.86\% and 9.33\%, and an average joint distance error of 0.193 and 1.336 pixels in Zoom and Meet, respectively, across all considered network settings. The maximum data rate corresponding to the reported SER values is 1.5 Kbps for Meet and 3 Kbps for Zoom. We further evaluate our joint prediction, error detection, and error correction mechanisms using existing ASL datasets. Our joint prediction method achieves a Mean Squared Error (MSE) of 0.1888 while being robust to noise up to a ten-pixel displacement. The proposed error detection mechanism identifies erroneous poses with an accuracy of 86.6\%.

We also assess the end-to-end performance of {\system} under two machine recognition tasks: (i) ASL alphabet recognition and (ii) ASL sign recognition under varying network conditions, using existing datasets. Our findings indicate that poses transmitted via \system achieve an average accuracy of 96.2\% and 95.75\% for ASL alphabet recognition in Zoom and Meet, respectively. For ASL sign recognition, poses transmitted over Zoom achieve an average accuracy of 88.64\%. Lastly, we conducted user studies with ASL speakers to evaluate the effectiveness of \system for ASL communication. We find that ASL speakers could comprehend the ASL pose videos, and the user-provided ASL glosses had an average semantic similarity score of 0.69. Our evaluation results indicate that \system can enable ASL communication between ASL speakers under lower bandwidth conditions where existing video communication technologies fail.

\section{Background and Motivation}
Modern customer-facing video conferencing applications (VCAs) are proprietary, with limited information available on their underlying design and architecture. However, the consensus is that they utilize some form of Real-time Transport Protocol (RTP) to transport media \cite{rtp1,rtp2} and use separate streams for audio and video \cite{measurement_IMC}. VCAs have multiple components that provide functionalities like media compression, bit rate adaptation, congestion detection, and multi-user streaming \cite{measurement_IMC}. Numerous studies have demonstrated the impact of network speeds on video quality in VCAs such as Zoom, Meet, Webex, and Teams \cite{measurement_IMC,measurement_1}.  The general findings indicate that network utilization varies from 0.8 Mbps to 1.9 Mbps, and the recovery time under lower speeds can be as significant as 25 seconds \cite{measurement_IMC}. The video quality measures, such as frames per second, freeze duration (a freeze in the received video), and quantization parameter (which controls the compression rate and resulting video quality), are varied based on the available network capacity \cite{measurement_1, measurement_IMC, dont_forget_user}. Although recent works have started investigating the impact of these parameters on end-user engagement \cite{dont_forget_user}, the effect on ASL communication is less understood. \textit{ We note that lessons learned from measurement studies that use talking heads to emulate a general video conferencing setting do not accurately represent the typical video dynamics observed in American Sign Language (ASL) communication.}

\begin{figure}
	\centering
	\vspace{-15pt}
		\includegraphics[width=0.5\textwidth]{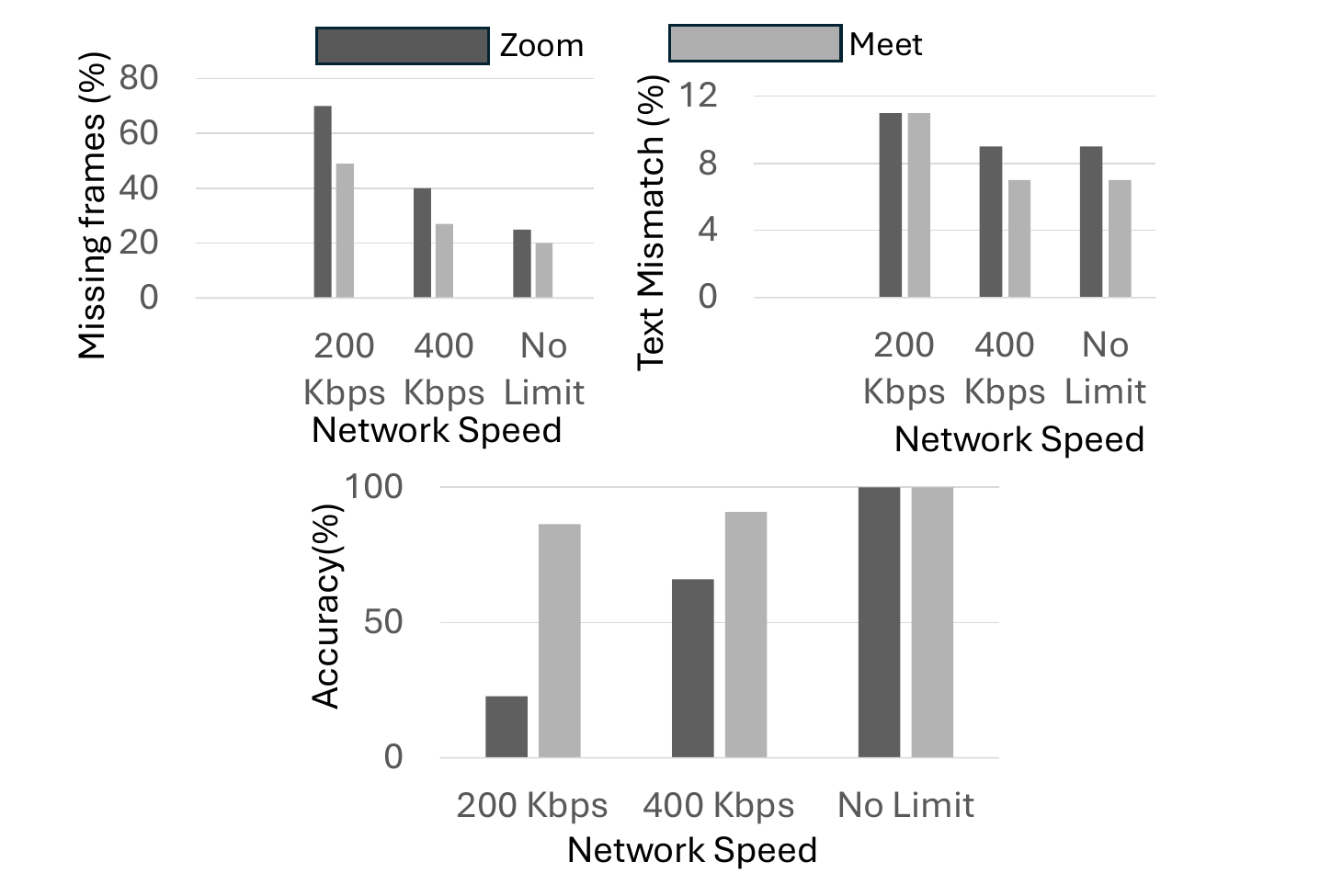}%
	\vspace{-15pt}
	\caption{Impact of network conditions on different video conferencing applications}
	\vspace{-15pt}
	\label{fig:missing_frames}
\end{figure}ASL is a visual-spatial language with its own grammatical rules. The language is expressed through the movements of the hand, body, and facial expressions \cite{ASL_1} and used for everyday communication by nearly five hundred thousand \cite{ASL_2, ASL_3} individuals. 
To examine the impact of network speed on ASL communication using various VCA, we conducted a pilot study. For this study, we simulate a video conferencing setup between two hosts and use virtual cameras \cite{virutal_cam} and virtual microphones \cite{VB_cable} to control the data fed into the VCA. For the video, we use ASL videos for studying ASL sign recognition \cite{WLASL}, and for audio, we use conversational audio between two speakers. We use NetLimiter \cite{netlimiterNetLimiter}, a popular network management tool, to control the network speed and vary it from 0.20 Mbps to 0.10 Mbps; these speeds are chosen based on recommended speeds by various VCA applications and prior measurement studies \cite{measurement_1,measurement_IMC,measurement_TON,measurement_webex}. The study focuses on answering two questions: (i) What is the impact of the network speed on ASL recognition? and (ii) Is there a difference between video and audio performance? Although the answer to the latter question is obvious based on the lower speed recommendations by VCAs for audio \cite{measurement_TON}, we still want to establish the exact difference. We randomly pick ten words from the world-level ASL recognition dataset \cite{WLASL} and retrain the proposed models to infer the impact on ASL recognition. To compare the effects on video and audio, we prelabel the transmitted frames and compare the missing frames in the received video and the missing words from the received audio by transcribing with state-of-the-art speech-to-text tools \cite{otterOtterMeeting}.

Fig.~\ref{fig:missing_frames} shows the results for word-level ASL recognition across two VCA applications under different network conditions. There is a stark decrease in the accuracy of the ASL recognition compared to the no network limit scenario as the network speed decreases across applications. However, there is also a difference in performance between these applications, which we attribute to the strategies adopted by each VCA in reaction to a change in network capacity. This difference is also evident in the number of missing frames across different applications, as seen in Fig.~\ref{fig:missing_frames}. While Meet significantly reduce video quality by adjusting the resolution and quantization parameters, Zoom maintains video quality while allowing for a longer freeze duration, where the last received frame is replayed, resulting in significant stalls on the receiver end. The difference in video quality degradation across various applications is also supported by other existing works \cite{measurement_IMC, measurement_TON}. Finally, the received audio quality is consistent across all three applications under varying network conditions, indicated by the percentage of correctly transcribed words and supported by existing measurement studies \cite{measurement_TON}.

	The major takeaways from our pilot studies are that the video optimization techniques used by current VCA’s can affect ASL communication under lower network speeds and that the differential network requirements for audio and video impact the receive quality differently. \textit{Building upon these observations, we propose a solution designed to facilitate ASL communication under network conditions that are not conducive to video transmission in existing VCAs.} Our key insight is that, for ASL speakers who do not use audio, the audio stream can serve as a reliable channel for communication during unfavorable network conditions. However, the transmitted information should convey visual spatial signifiers, such as hand shape and position, that are key to ASL, while maintaining a lower network speed requirement. The human pose information \cite{openpose,mediapipe}, which comprises the human skeleton joints, referred to as pose keypoints in the literature, estimated from an image \cite{openpose,mediapipe}, is a viable candidate. The human pose is orders of magnitude smaller in size compared to the image; the raw size of an image of resolution $720 \times 1280$ is approximately $2.75 MB$ while the equivalent pose containing 137 key points \cite{mediapipe} stored as sixteen-bit floating point numbers is $0.0005226 MB$. The human pose has been established as a stable representation for American Sign Language (ASL) recognition \cite{ASL_rep} and has been utilized by a range of existing works for ASL sign recognition \cite{ASL_recognize}, ASL synthesis \cite{ASL_generate}, and ASL anonymization \cite{ASL_anonymize1, ASL_anonymize2}. 

\section{System Overview}
\label{sec:system}
\begin{figure}[t]
	\vspace{-15pt}
	\centering
		\includegraphics[width=0.5\textwidth]{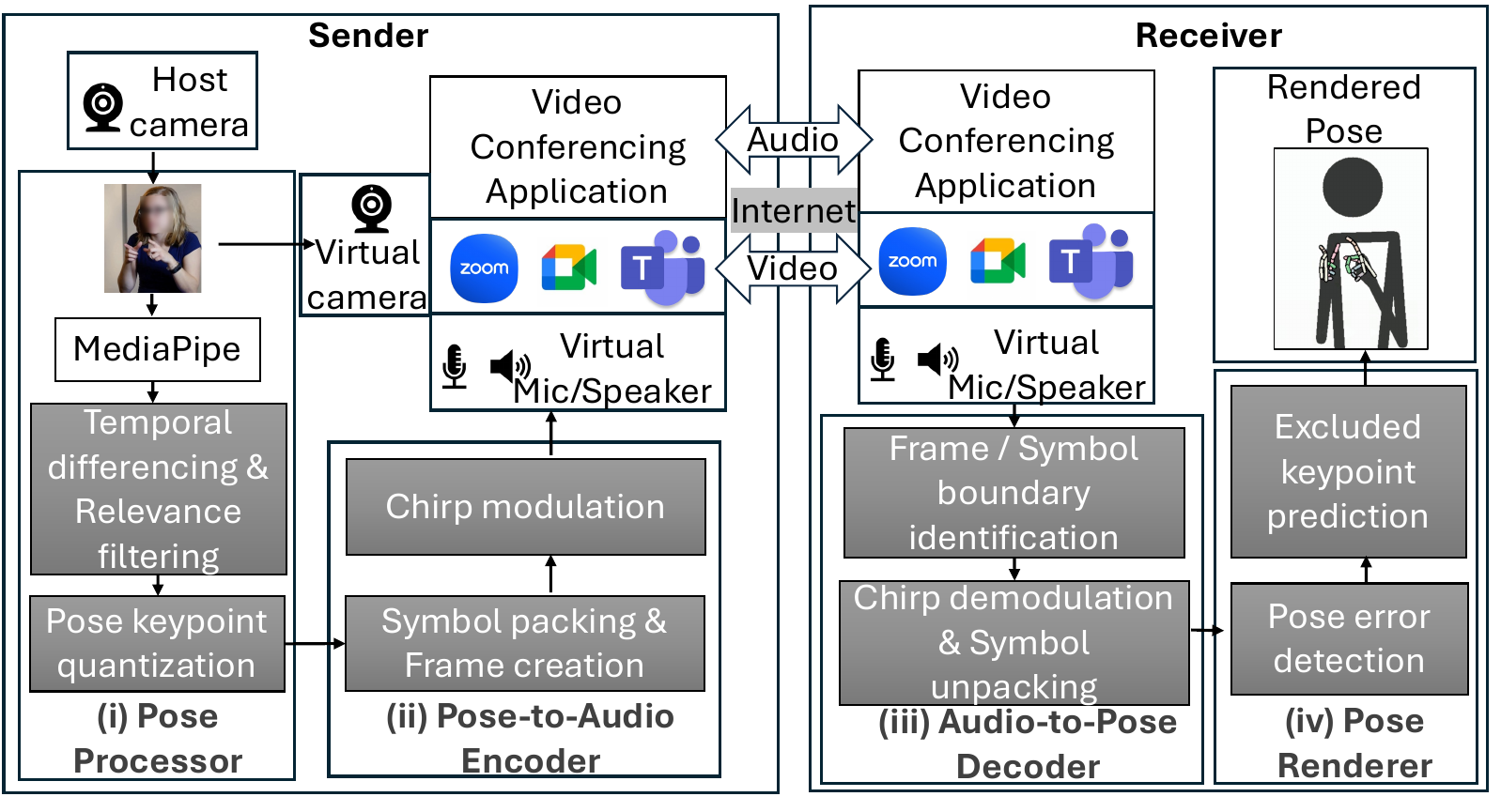}
		
	\caption{{\system}'s different components.}
		\vspace{-10pt}
        \label{fig:system_overview}
\end{figure}
Fig.~\ref{fig:system_overview} shows the complete end-to-end pipeline for \system. We assume a one-to-one video conferencing setup between two American Sign Language (ASL) speakers; our implementation can also be extended for multi-speaker and ASL speaker-interpreter setups. We use virtual cameras \cite{virutal_cam} and microphones \cite{VB_cable} to mediate the interaction between \system and Video Conferencing Applications (VCAs). 

\subsection{Sender side:}
The sender side comprises two components: 	
\begin{itemize}
    \item \textbf{(i) Pose Processor:} Directly interfaces with the host camera, captures the image, sends a copy to the virtual camera connected to the VCA, and another to the MediaPipe for body and hand pose estimation. The extracted pose information is analyzed for possible reduction in size through temporal differencing and visibility/relevance filtering. The final subset of selected pose keypoints are quantized and passed to the next component.
\item \textbf{(ii) Pose-to-Audio Encoder:} The quantized pose keypoints are packed into symbols. The symbols from the same image are grouped into a single frame, prefixed with a frame identifier, and modulated into chirp signals, which are then input to the virtual microphone connected to the VCA.
\end{itemize}
\subsection{Receiver side:}
	The receiver side comprises two components:
    \begin{itemize}
\item \textbf{(iii) Audio-to-Pose Decoder:} Interfaces with a virtual speaker/microphone connected to the VCA and captures audio. The captured audio is processed for audio frame identification, precise chirp symbol boundary estimation, and segmentation. The segmented chirp symbols are demodulated, and the resulting symbols are unpacked and grouped to obtain the quantized pose values.
\item \textbf{(iv) Pose Renderer:} The pose renderer detects and corrects errors in received pose values and predicts the remaining excluded pose keypoints. The predicted and transmitted pose keypoints are combined to create the final pose image, which is then displayed to the user.
\end{itemize}

\section{Pose Processor:}
\begin{figure}[t]
	\centering
	\vspace*{-10pt}
	
		\includegraphics[width=0.4\textwidth]{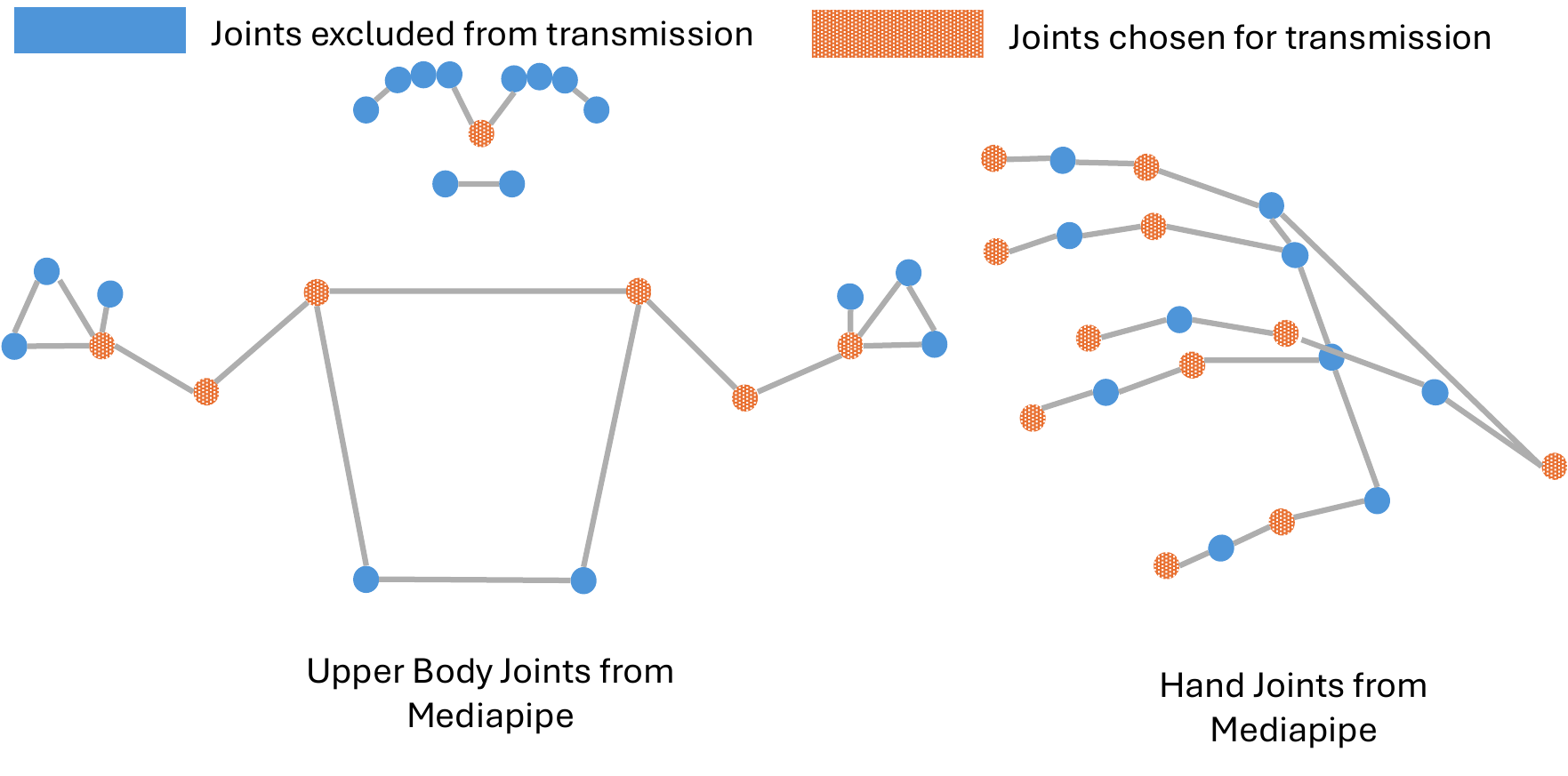}
			
	\vspace{-5pt}
	\caption{Upper body and hand keypoints provided by Mediapipe and the joints transmitted by \system}
		\vspace{-10pt}
        \label{fig:mediappipe_chosen_joints}
\end{figure}
MediaPipe provides twenty-one hand pose keypoints and thirty-three body pose keypoints \cite{mediapipe_pose,mediapipe_hand}. Although pose estimation solutions are generally robust, they still perform poorly in blurry and low-quality images, resulting in incorrect and missing pose keypoint estimates \cite{how2sign}. Additionally, the need to operate under extreme network conditions and tolerate the impact of the codec reduces the data capacity, as detailed in Section.~\ref{sec:codec-impact}. We propose two approaches to address these challenges: first, we track the displacement in pose keypoints over subsequent images, and if the displacement is insignificant, we only transmit the displacement. Second, we filter out pose keypoints that are invisible in the image, select a subset of them, and quantize these keypoints to a lower resolution. 

Let $P_t$  and  $P_{t-k}$ be the set of 2D pose keypoints obtained at time $t$ and $t-k$. We compute the per joint displacement between the two sets, and if the displacement values for all the joints are less than the threshold $\delta$, we only send the displacement. The threshold $\delta$ is decided based on the bits per symbol. We retain a prior pose obtained at time $t-k < t$ for this. Whenever the keypoint displacements are greater than the threshold $\delta$, the pose keypoints $P_t$ will replace $P_{t-k}$, and the pose key points are sent. Sending the displacement reduces the number of symbols by half and helps to maintain a higher poses-per-second ($P_s$)  rate. If displacements are to be sent, the values are passed directly to the Pose-to-Audio Encoding Engine without requiring further processing. If not, we filter pose keypoints based on visibility and select a subset of the upper body and hand keypoints. Fig.~\ref{fig:mediappipe_chosen_joints} shows the chosen keypoints from the complete set. Some of the keypoints in the body, such as the fingers, are skipped due to redundancy with the hand estimates. The retained hand keypoints are sufficient to interpolate for those that are missing on the receiver side. Following this, we quantize the retained 32 pose keypoints. The output from MediaPipe is a 16-bit floating-point number in the range zero to one, normalized to the width and height of the image. Sending the keypoints as a 16-bit representation results in $128$ Bytes data for a single image; we reduce this to  $56$ Bytes by quantizing them to seven-bit integers.

\section{Pose-to-Audio Encoder:}
	The quantized pose keypoints should be converted into audio and input into the video conferencing application (VCA), which transmits them over the network and receives them on the receiver end. A straightforward solution is to first convert the pose keypoints into bits (0 and 1) and use established modulation techniques to encode the generated bits into the audio, then input them into the VCA. In essence, our solution involves establishing a communication channel using audio as the signal, which is transmitted over the internet via the VCA. Our solution has to withstand two key challenges:
    \begin{itemize}
	\item Most VCAs use proprietary configurations to compress audio, and the general practice is to adapt the compression rate in response to the changes in network conditions. Our proposed modulation scheme should be robust to these rate adaptations, even without knowledge of the underlying system operations.
	\item As the data traverses through the internet, there could be a loss of information at the network or link level, and our solution should be tolerant to such losses.
    \end{itemize}
Let $C(\cdot)$ represent the codec that compresses the audio, $N(\cdot)$ represent the network-level losses encountered by the compressed audio, and $x(t)$ represent modulated pose key point audio signal sent from the sender side. The audio received on the receiver side is a function of $C(\cdot)$ and $N(\cdot)$
\begin{equation}
    r(t) = N(C(x(t)))
\end{equation}
Although we represent the impact of the codec and the network separately, in practice, the codec parameters are adapted in response to changing network conditions. 
\subsection{Audio Codecs in VCA} 
To study the codec's impact on modulated audio, we examine the Opus codec \cite{opus}, used in major VCAs such as Zoom and Meet \cite{Meet_opus,Zoom_opus}. Opus operates in two modes: SILK and Constrained Energy Lapped Transform (CELT). SILK encodes speech using Linear Predictive Coding (LPC) with a maximum sampling rate of 24 kHz \cite{opus_codec_paper}. LPC employs a predictive model to estimate future samples from past ones, reducing transmitted information \cite{LPC}. CELT encodes music with transform coding, supporting sampling rates up to 48 kHz. Transform coding splits signals into overlapping frames, converts them to the frequency domain, quantizes coefficients, and transmits them. Opus uses the Modified Discrete Cosine Transform (MDCT) for this step.

During audio encoding, latency and quality are mainly controlled by the frame size ($f$) and bitrate ($b$). Smaller $f$ reduces latency but lowers quality, while lower $b$ decreases both quality and size. The Opus codec supports $f \in [2.5, 60]$ ms and $b \in [6, 256]$ Kbps, with defaults of $f=20 $ ms and $b=64 $Kbps at sampling rates above 44 kHz \cite{opus,opus_mac}. VCAs typically use variable bitrate (VBR), adapting $b$ and $f$ dynamically to network conditions. With VBR, the encoder operates on frames indexed by $m$, with frame duration $f_m$ ms and bitrate $b_m$ Kbps that can change per frame. Let the codec operator for frame $m$ be $C_{b_m,f_m}$. If $x_m(t)$ is the modulated audio restricted to frame $m$ (i.e., $t\in[\tau_m,\tau_{m+1}$), then the received signal is
\begin{equation}
r_m(t) \;=\; N\!\big(C_{b_m,f_m}\{x_m(t)\}\big),\qquad t\in[\tau_m,\tau_{m+1})
\end{equation}
Equivalently, in continuous time with time-varying parameters,
\begin{equation}
r(t) \;=\; N\!\Big(C_{\,b(t),\,f(t)}\{x(t)\}\Big),
\end{equation}
where $b(t)$ and $f(t)$ are functions that follow the frame schedule $\{\tau_m\}$. \textit{We need a modulation scheme that is robust to these time-varying adaptations to $b$ and $f$.}
\begin{figure*}[hb]
	\vspace{-15pt}
	\centering
    \captionsetup[subfloat]{labelfont=scriptsize}

    \subfloat[]{
		\includegraphics[width=0.4\textwidth]{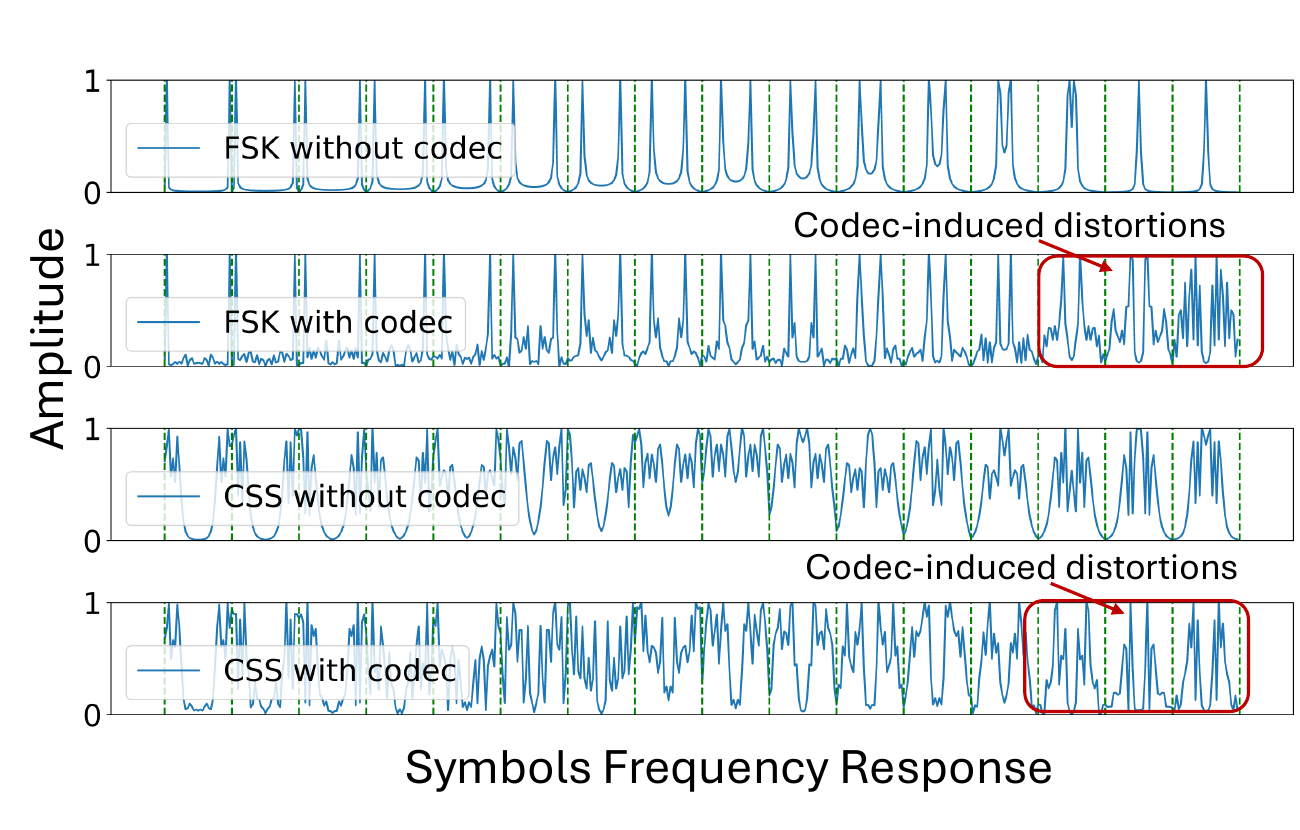}
		\label{fig:symbols}
	} 
    \hspace*{-10pt}
    	\subfloat[]{
		\includegraphics[width=0.3\textwidth]{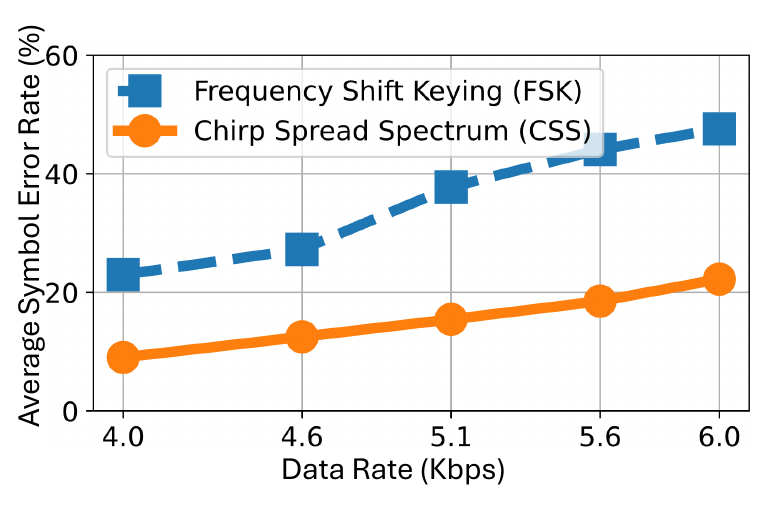}
		\label{fig:data_rate_compare}

	} 
    \hspace*{-10pt}
	\subfloat[]{
		\includegraphics[width=0.3\textwidth]{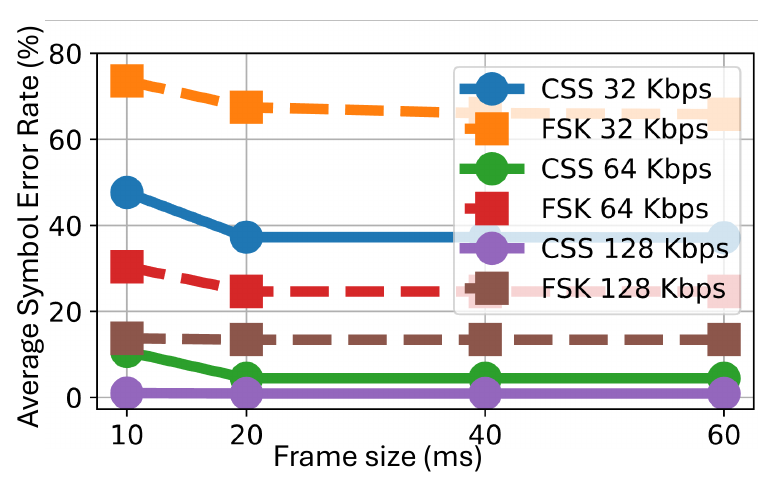}
		\label{fig:codec_config_compare}
	} 
	\vspace{-5pt}
	\caption{Comparing Frequency Shift Keying (FSK) and Chirp Spread Spectrum  (CSS). (a) Different FSK and CSS symbols frequency response with and without codec compression. (b) The impact of data rate on the average SER averaged for different codec configurations. (c) The impact of codec configurations on the average SER averaged for different data rates.}
		\vspace{-10pt}
\end{figure*}

\subsection{Designing a Robust Modulation Scheme} 
Prior work in audio steganography has extensively examined the challenges posed by audio codecs when embedding information in audio and proposed methods to mitigate distortions \cite {audio_steg_survey,audio_seg_review2}. Most approaches assume full knowledge of the codec, and some even modify it. Least Significant Bit (LSB) encoding \cite{lsb,lsb1,lsb2} is simple but fragile against compression \cite{audio_steg_survey,audio_seg_review2}. Amplitude- and phase-based techniques depend on preserving magnitude and timing, yet codecs alter both; automatic gain control disrupts amplitude encoding, while transform coding discards phase information \cite{opus_codec_paper,opus_codec2}. Echo cancellation and frame segmentation further degrade fidelity, making these methods unreliable for transmitting symbols through VCA audio channels. In contrast, frequency modulation techniques are far more robust and can withstand codec-induced signal alterations.

\subsubsection{Codec's impact on different modulation techniques}
\label{sec:codec-impact}
Frequency modulation schemes such as Frequency Shift Keying (FSK) encode bits using tones (e.g., $f_0$ for zero, $f_1$ for one), with each tone treated as a symbol \cite{fsk,wifi_symbols}. Chirp Spread Spectrum (CSS) instead uses frequency sweeps, e.g., $f_0 \to f_1$ for zero and $f_1 \to f_0$ for one \cite{css}. CSS, widely used in long-range wireless systems, is more robust to low SNR and frequency-domain distortions from transform coding than FSK’s single-tone symbols \cite{lora_tutorial}. Beyond robustness, higher data rates are needed to increase poses-per-second ($P_s$) and improve user experience. The rate is $R_b = R_s \log_2(M),$ where $R_s$ is the symbol rate and $M$ the modulation order. Increasing $R_s$ requires reducing samples per symbol ($N_s$) under a fixed sampling rate ($f_s$, capped at 48 kHz in VCAs), while ensuring Nyquist’s condition $N_s > \tfrac{1}{2}\max(f_0,f_{M-1})$
\cite{nyquist}. To compare FSK and CSS, we implemented both, generated symbols, and passed them through the Opus codec \cite{opus_mac}, measuring symbol error rate (SER) after encoding, decoding, and demodulation. We varied $R_b$ by adjusting $M$ and $N_s$, and tested codec effects using frame sizes of 10–60 ms and bitrates of 32–128 Kbps, following Opus recommendations.

Fig.~\ref{fig:symbols} shows a subset of the symbols' frequency response with and without codec compression with the default codec settings from Zoom. The FSK symbols at higher frequencies undergo significant distortion under compression; however, the CSS symbols undergo less distortion as they spread the symbols across frequencies, making they more resilient to the codec compression. Fig.~\ref{fig:data_rate_compare} shows the average Symbol Error Rate (SER) for the two modulation schemes for different data rates averaged for chosen frame size and bit rate configurations. The average SER gap between Frequency Shift Keying (FSK) and Chirp Spread Spectrum (CSS) modulation schemes starts at ten percent and reaches as high as twenty-five percent for the highest data rate configuration ($6$Kbps). The increase in data rate is achieved by decreasing the number of samples per symbol ($N_s$), which increases the overlap between symbols during codec compression, resulting in information loss and an increase in errors during the demodulation of the decoded audio. Although CSS suffers performance loss in high data rate scenarios, the loss is significantly lower compared to FSK. Fig.~\ref{fig:codec_config_compare} provides further detail on the impact of codec configurations on the two modulation schemes. The most challenging configurations are those with a frame size less than twenty and bit rates less than sixty-four. These settings optimize for the lowest latency and smallest compression size possible, which explains the higher information loss and higher average SER. These configurations represent extreme scenarios, and the suggested default values are $20$ ms and $64$ Kbps for the frame size and the bit rate \cite{opus_mac}. \textit{The results indicate that the CSS modulation scheme is resilient to various codec configurations and offers superior performance. Ultimately, spreading each symbol over a range of frequencies gives resilience to the compression techniques that segment and transform audio signals during compression. }

\begin{figure}[h]
	\vspace{-5pt}
	\centering
		\includegraphics[width=0.5\textwidth]{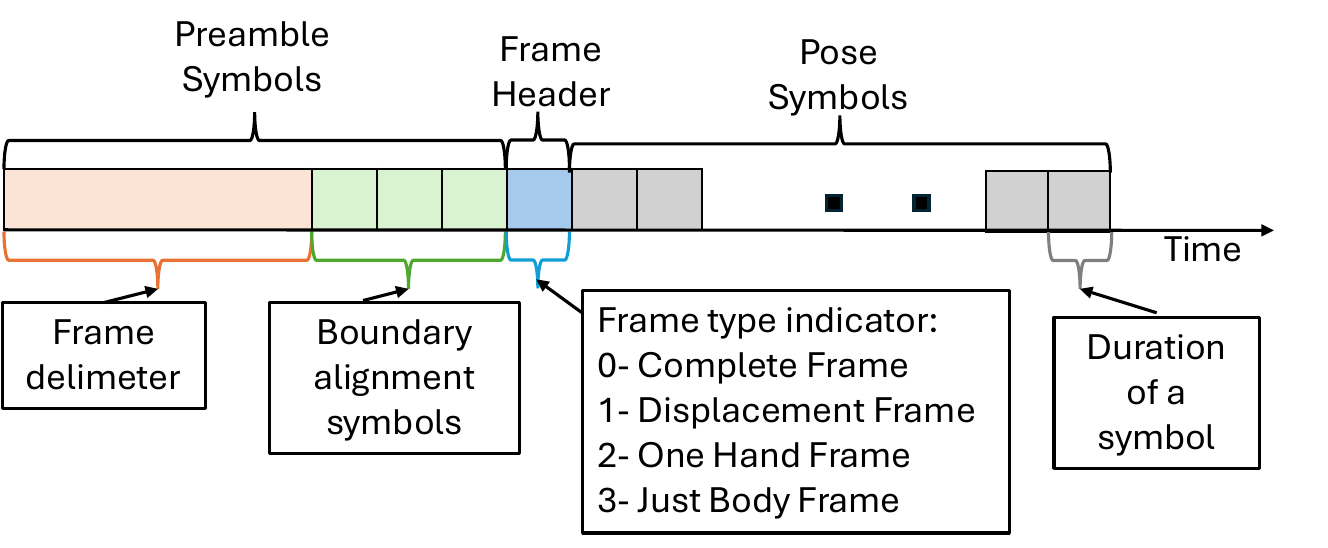}
		
	\caption{{\system}'s frame design showing the different parts of the frame and their durations.}
		\vspace{-10pt}
        \label{fig:system_frame}
\end{figure}

\subsubsection{{\system}’s Modulation Scheme and Frame Design}
	We use Chirp Spread Spectrum (CSS) for modulation, setting the modulation order ($M$) and the symbol rate ($R_s$) to 16 and 1500 symbols/s, respectively. The chosen configuration bounds the data rate ($R_b$) capacity of our scheme to $6 $ Kbps. The quantized pose keypoints obtained from the Pose Processing Engine result in 56 Bytes, which, with our modulation scheme of four bits per symbol, yields 112 symbols. If we send the pose symbols continuously, we can achieve a $\approx 13$ poses-per-second rate. However, to render the pose keypoints, we need to separate the pose keypoints of each image, and accurate demodulation requires the precise location of symbol boundaries. We address this by grouping the symbols corresponding to a single image into a frame and using preambles to identify the beginning of each frame. For the frame delimiter, we use a $1 $ kHz sinusoid of duration $12.7$ ms. We tested different delimiters, including those used in LoRa \cite{lora_wiki,lora_tutorial}, and found that the $1 $ kHz frame delimiter at a longer symbol duration offers accurate and low-latency frame detection.

Fig.~\ref{fig:system_frame} shows a single frame with the duration for different parts of the frame in terms of the symbol duration ($T_s$), which for the proposed scheme is $0.667$ ms. Following the frame detector, we use three up chirp symbols ($f_0 \to f_1$) for accurate symbol boundary detection. A single symbol is used as a frame header, indicating both the type of the frame and its duration.  Currently, \system supports four types of frames: 
\begin{itemize}
\item Complete frames: All 112 body and hand pose symbols are sent. 
\item Displacement frames: We only send the difference between consecutive poses, resulting in the shortest frame duration. 
\item One hand frames: When single-handed signs are performed, only the body and a single hand's pose symbols are sent.
\item Just body frame: When both hands are missing in an image, we only send the body pose symbols.
\end{itemize}

To keep the design simple, we accommodate only four frame sizes, corresponding to the four frame types. Knowing the frame type allows the system to estimate the next frame start and the expected count for the received pose symbols. \textit{In summary, {\system}’s modulator receives quantized pose symbols, modulates them, and packs them into a frame with headers and preambles in place and passes it to the video conferencing applications' audio stream.}

\section{Audio-to-Pose Decoder}

The transmitted audio through the VCA is received on the receiver side via the virtual microphone and speaker interfaces over the internet. Before demodulating the audio to recover the pose symbols, it is necessary to accurately identify the beginning of each frame and segment the stream into individual symbols. This step is critical to the \system, as false positives in frame detection can produce incorrect pose frames, while false negatives can result in the loss of pose frames. Likewise, inaccuracies in symbol boundary identification can lead to symbol errors during demodulation. Overall, the use of less effective schemes for frame and symbol boundary detection may significantly degrade the quality of the recovered poses, hindering ASL communication and negatively impacting the user experience.

\subsection{Frame and Symbol boundary Identification}
\system continuously captures the received audio and searches for the frame delimiter. As this process occurs every few milliseconds, depending on the frame duration, a simple matched filter is employed to identify the beginning of the frame. A matched filter is commonly used to detect a known signal or template within an unknown signal \cite{matched_filter_wiki}. In this context, the $1\text{kHz}$ frame delimiter corresponds to the known signal $s(t)$, and the received audio is denoted as $r(t)$. The matched filter is defined as the time-reversed version of the known signal. The filtered output $y(t)$ is obtained through the convolution between the received signal and the matched filter:

\begin{equation}
y(t) = (r * h)(t),
\end{equation}

where the estimated frame start time $\hat{t}_0$ is defined as

\begin{equation}
\hat{t}_0 = \arg\max_t y(t).
\end{equation}

Frame detection is considered successful if

\begin{equation}
y(\hat{t}_0) \geq \gamma,
\end{equation}

where $\gamma$ is a predefined detection threshold. 
\begin{figure}[t]
	\vspace{-5pt}
	\centering
		\includegraphics[width=0.5\textwidth]{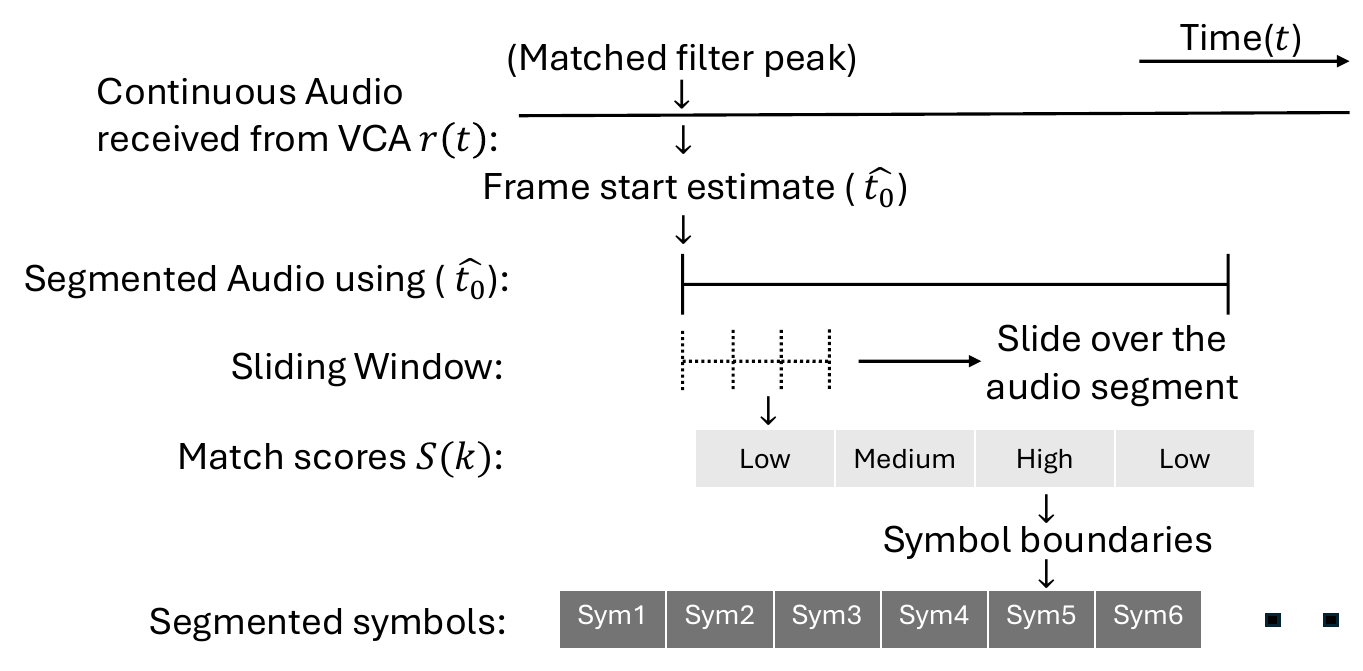}
		
	\caption{{\system}'s proposed approach for frame start identification and precise symbol boundary estimation.}
		\vspace{-10pt}
        \label{fig:frame_synchronization}
\end{figure}
Although the initial frame start estimate obtained from the matched filter is computationally efficient, it lacks the precision necessary for accurately determining symbol boundaries. Relying solely on this coarse estimate can result in significant segmentation errors. To address this, during transmission, we appended three up-chirp symbols following the frame delimiter, constituting the frame preamble. These known symbols enable fine synchronization, a technique commonly adopted in wireless communication systems \cite{wireless_fundamentals}. Starting from the coarse frame estimate $\hat{t}_0$, we extract a segment of the received audio and slide a window of size $N_s$ (the number of samples per symbol) across this segment. At each window position, we evaluate the similarity between the received samples and the expected three up-chirp symbols. The window that yields the maximum similarity score is selected as the precise frame start for symbol segmentation. To compute the match score at each window position, we demodulate the samples corresponding to the three known symbols, perform a Fast Fourier Transform (FFT) on each segment, and sum the magnitudes at the expected frequency bins (corresponding to the zero-bin of up-chirps). The total summed magnitude across the three segments is used as the matching metric.

Let $k$ be the current offset of the sliding window from the current frame start estimate. Then, the match score, $\mathcal{S}(k)$, from this offset is defined as 
\begin{equation}
\mathcal{S}(k) = \sum_{m=0}^{2} \left| X_m[k](f_0) \right|
\end{equation}

Where $m \in \{0,1,2\}$ indexes the three expected up-chirp symbols and $X_m[k](f_0)$ is the magnitude of the FFT of the $m$-th windowed segment at frequency bin $f_0$,    $ f_0 $ corresponds to the zero-bin (lowest frequency bin) under our configuration. Fig.~\ref{fig:frame_synchronization} illustrates the sequence of steps involved in obtaining segmented symbols from the received audio.

\subsection{Demodulation and symbol unpacking} The segmented symbols are first demodulated and unpacked into bytes, after which the values corresponding to the hands and body are separated into their respective groups. For Chirp Spread Spectrum (CSS) modulation, demodulation proceeds through a two-step process. Since each symbol is a cyclically shifted version of the reference up-chirp, the received chirp is initially dechirped by multiplying it with the conjugate of the reference up-chirp \cite{lora_demodulation}. As our received audio signals lack phase information, dechirping transforms into a product with the up-chirp. Subsequently, a Fast Fourier Transform (FFT) is applied to extract the original symbol value. The recovered symbols are then unpacked into bits and grouped appropriately to reconstruct the quantized pose keypoints that were transmitted. Once the keypoints are recovered, they undergo further processing for error correction and the prediction of any excluded keypoints. The final set of corrected pose keypoints is then used to render a complete pose frame for display to the ASL speaker.

\section{Pose Renderer}

\begin{figure}[t]
	\vspace{-15pt}
	\centering
	\hspace*{-10pt}
        \captionsetup[subfloat]{labelfont=scriptsize}
	\subfloat[]{
		\includegraphics[width=0.23\textwidth]{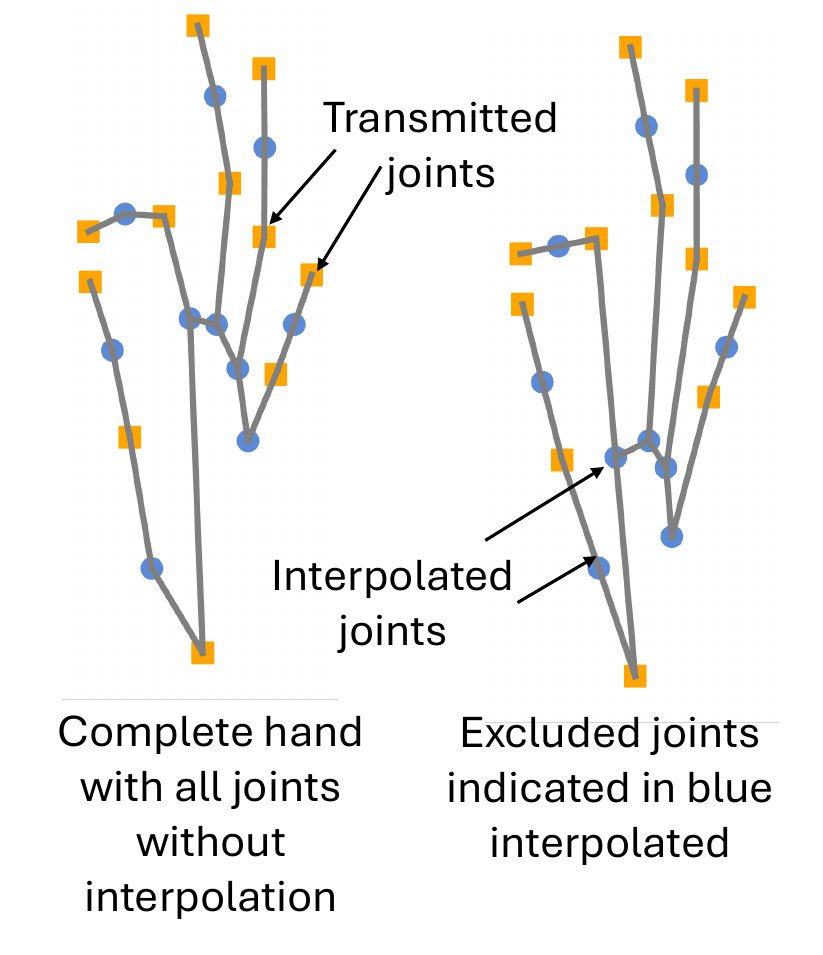}
		\label{fig:interpolation_example}

	} 
    		\hspace*{-15pt}
	\subfloat[]{
		\includegraphics[width=0.29\textwidth]{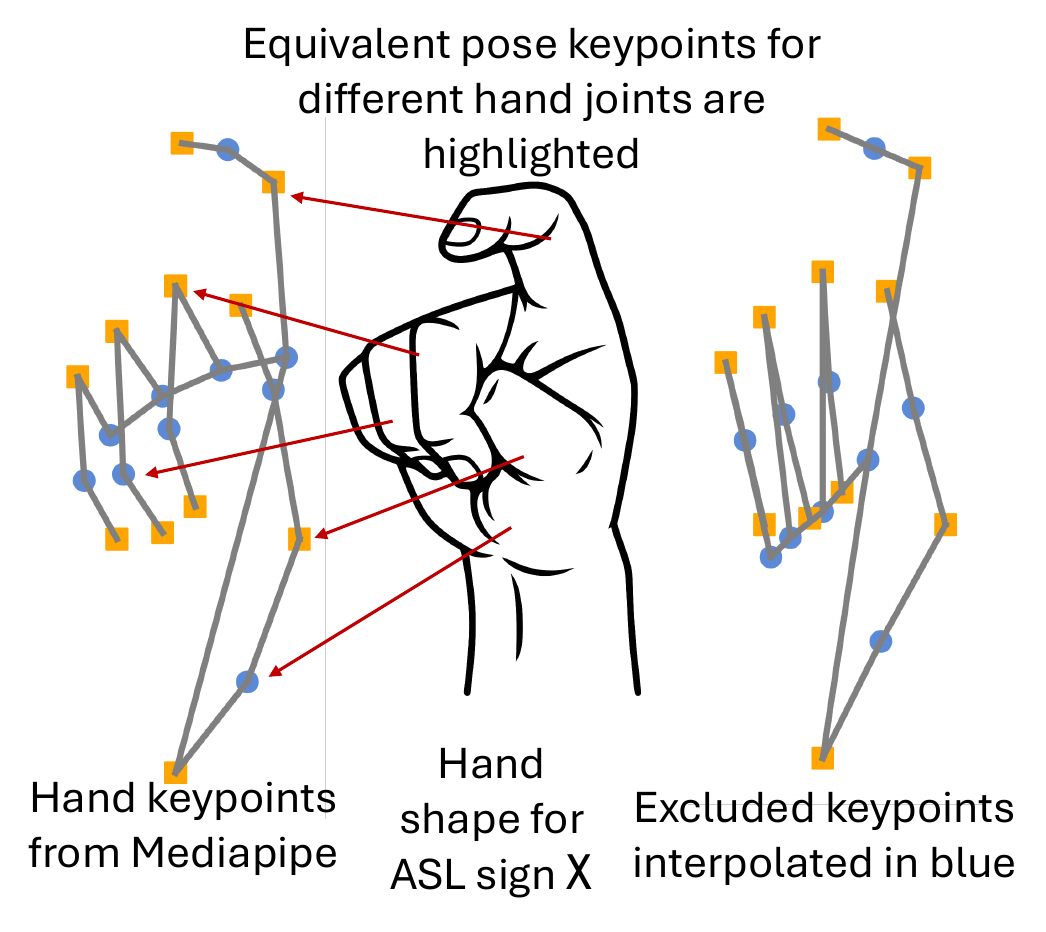}
		\label{fig:interpolation_error}
	} 
	\vspace{-5pt}
	\caption{Leveraging structural relationships between hand pose keypoints: (a) Structural relations between hand pose key points can predict hand pose keypoints, excluded from transmission. (b) Linear interpolation to predict missing keypoints does not preserve the structure of the original pose keypoints, leading to visual ambiguity.}
		\vspace{-10pt}
\end{figure}
As demonstrated in Section~\ref{sec:codec-impact}, audio codecs introduce a non-trivial impact on the symbol error rate (SER), potentially leading to errors in the reconstructed pose keypoint values. In addition to codec-induced errors, errors arising from adverse network conditions must also be anticipated. Since our objective is to enable communication under poor network conditions, the presence of such errors should be considered the norm rather than an exception. In network-based applications, forward error correction (FEC) and related coding techniques are commonly employed to detect and correct bit errors \cite{wikipediaErrorCorrection}. However, these mechanisms inherently consume additional data rate, which imposes constraints on their applicability in our low-data-rate setting. Furthermore, to maintain a higher poses-per-second rate, we reduced the number of pose keypoints transmitted from the sender. While necessary for efficiency, displaying only a subset of keypoints may degrade the user experience and diminish the effectiveness of communication.

Although the challenges are significant and common to any network-based application, \system possesses a unique advantage that others lack. Unlike general networked applications where bit errors irreversibly corrupt transmitted data, \system benefits from the inherent spatial and temporal structure of human poses. The connectivity among pose keypoints forms a graph-like structure, a principle widely leveraged in pose estimation and action recognition \cite{openpose, graph_pose1, graph_pose2}. This structural redundancy enables receive-side error correction. For example, Fig.~\ref{fig:interpolation_example} illustrates how simple linear interpolation can reconstruct untransmitted hand keypoints, preserving the structural similarity to the original hand. We propose approaches that leverage the intrinsic relationships among pose keypoints to correct transmission errors and predict omitted joints.

\subsection{Excluded Keypoint prediction}
\label{sec:point-detection}
Hand shapes are crucial for ASL communication, and two ASL signs with similar movement patterns can often be distinguished only by hand shape \cite{asl_green, asl_blue}. Furthermore, for uncommon nouns, it is typical to use fingerspelling, where each letter of the alphabet is represented by a unique hand shape \cite{asl_fingerspelling}. Simple linear interpolation of excluded keypoints fails to preserve the structural similarity critical for handshape recognition. Fig.~\ref{fig:interpolation_error} illustrates this for the ASL sign \texttt{X}. As shown, while keypoint estimates obtained from Mediapipe maintain visual similarity to the actual hand shape, the interpolated keypoints do not. This discrepancy arises because linear interpolation accounts only for relative joint distances, neglecting the relative angles between joints. The problem is further compounded by the fact that our joint estimates are in two dimensions, where camera viewing angle and distance can distort joint positions due to perspective projection \cite{perspective_projection}.

Finally, as \system operates in real-time, the proposed solution must maintain low latency. To address this constraint, we learn a function that takes the transmitted pose keypoints and predicts the excluded ones. Formally, let $P_t \in \mathbb{R}^{N \times d}$ denote the pose keypoints received at time $t$, where $N$ is the number of transmitted keypoints and $d$ is the dimensionality (e.g., $d = 2$ for 2D keypoints). We define a mapping $\mathcal{F}: \mathbb{R}^{N \times d} \rightarrow \mathbb{R}^{(M + N) \times d}$ that jointly predicts both the transmitted and excluded keypoints, where $M$ is the number of excluded joints. This formulation allows the model to learn structural dependencies among all keypoints and enables correction of potential errors in the transmitted subset. The final reconstructed pose at time $t$ is given by $\mathcal{F}(P_t)$. To simplify the learning objective, we restrict training to poses captured from a frontal view, assuming the user’s torso is approximately parallel to the camera’s image plane, an assumption consistent with typical video conferencing usage.

We use a Multi-Layer Perceptron (MLP) with two hidden layers to learn the mapping function. 
The model takes the transmitted hand keypoints as input and predicts the excluded pose keypoints. Each hidden layer contains 128 units, and the ReLU activation function \cite{relu} is used for the hidden layers. We use the How2Sign \cite{how2sign} dataset to train the model, retaining only frontal-view frames and extracting pose keypoints with Mediapipe. Our dataset consists of approximately 296,000 samples, with $80\%$ used for training and $20\%$ for testing. We optimize the model using the Adam optimizer \cite{adam_optimizer} with a learning rate of $0.001$ and a batch size of $100$. The training objective is to minimize the Mean Squared Error (MSE) \cite{MSE} between the predicted and ground-truth keypoints. We train separate models for the left and right hands and use the chosen body keypoints as such.

\subsection{Error Detection and Correction}
From the results in Section~\ref{sec:codec-impact}, we know that \system must tolerate at least $20\%$ symbol errors; however, we still need to investigate system's performance when the error rate exceeds this threshold. Error rates are expected to increase under fluctuating network conditions and as video conferencing applications (VCAs) adapt to maintain stability. Displaying all received pose frames without error detection or correction would inevitably degrade the user experience, a phenomenon commonly observed in many video platforms. To address this, we adopt a two-step strategy that leverages the structural relationships between pose keypoints in typical video conferencing setups. First, we detect and filter out outlier poses, and then apply error correction techniques.
\begin{figure}[t]
	\vspace{-15pt}
	\centering
        \captionsetup[subfloat]{labelfont=scriptsize}
	\hspace*{-10pt}
	\subfloat[]{
		\includegraphics[width=0.25\textwidth]{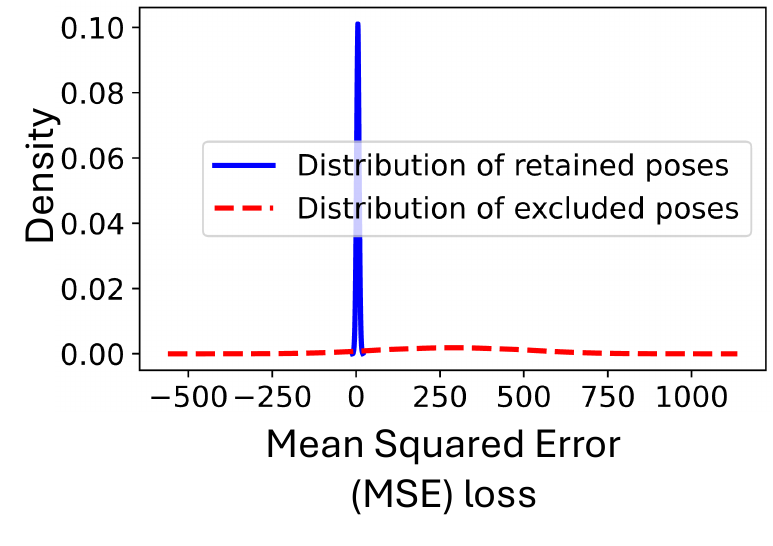}
		\label{fig:loss_distribution}

	} 
    		\hspace*{-10pt}
	\subfloat[]{
		\includegraphics[width=0.26\textwidth]{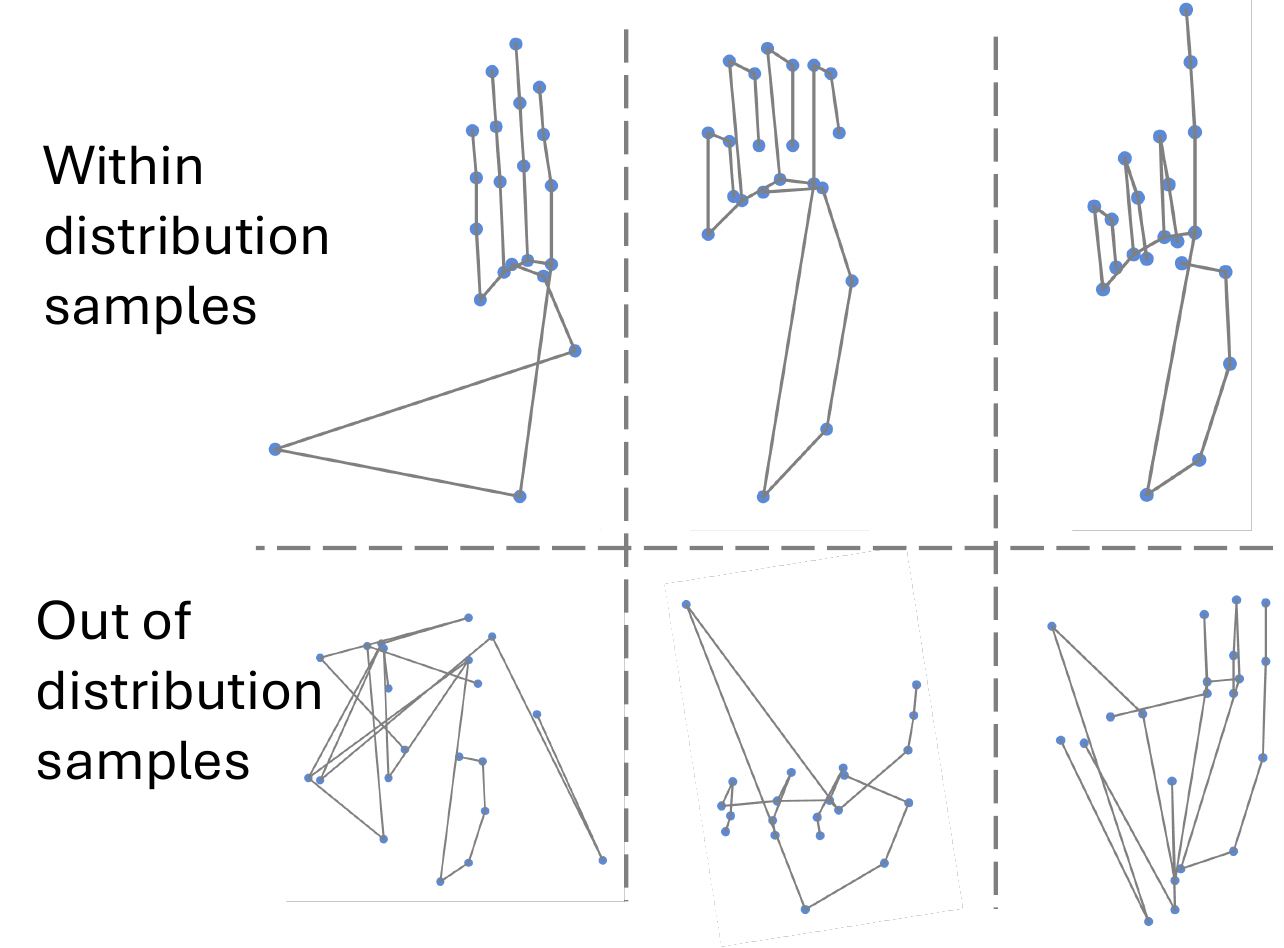}
		\label{fig:ood_samples}
	} 
	\vspace{-5pt}
	\caption{The different distributions and examples of samples which are retained for further processing and those that are discarded by \system. (a) Loss distribution of samples that are retained and those that are omitted (b) Examples of samples within distribution and out of distribution}
    \label{fig:ae_fig_example}
		\vspace{-10pt}
\end{figure}
\subsubsection{Error Detection}
We redefine the problem of error detection as an out-of-distribution (OOD) detection problem. Specifically, we propose to use autoencoders, a class of models widely explored for outlier detection and OOD sample identification \cite{auto_enc2,autoenc_1}. Autoencoders help circumvent the challenge of explicitly defining OOD samples. An autoencoder consists of two functions: an encoder $E(\cdot)$ and a decoder $D(\cdot)$. The encoder maps input samples $x \in \mathbb{R}^d$ to a lower-dimensional latent representation $z = E(x)$, where typically $\dim(z) \ll d$. The decoder attempts to reconstruct the original input from the latent space, yielding $\hat{x} = D(z)$. The latent features $z$ aim to preserve the underlying distributional structure of the input data. In particular, variational autoencoders (VAEs) model the latent space explicitly as a Gaussian distribution \cite{vae} and have been used for both sample generation and probabilistic modeling of data distributions \cite{pu2016variational}.

We use Multi-Layer Perceptrons (MLPs) for both the encoder and decoder, each comprising a single hidden layer with ReLU activation functions to introduce non-linearity \cite{relu}. The encoder takes as input all received pose keypoints for the hands and body and projects them into a latent space, while the decoder reconstructs the original pose keypoints from the latent representations. The hidden layer and latent space dimensions are set to $x$ and $y$, respectively. We train the model using the How2Sign dataset \cite{how2sign} described earlier, adopting the same training configuration, optimizer, and objective function as previously outlined.

Once trained, we use the reconstruction loss of a sample, computed as the mean squared error (MSE) between the input and the reconstructed output of the autoencoder, as a proxy for detecting out-of-distribution (OOD) samples. We define a loss threshold as a $20\%$ increase relative to the maximum reconstruction loss observed over the training samples. The underlying intuition is that the autoencoder learns to reconstruct samples consistent with the training distribution, which consists of clean poses. Consequently, when erroneous pose samples are input, the reconstruction loss is expected to be significantly higher than that for clean poses. The $20\%$ threshold reflects the error margin we aim to tolerate based on the expected network-induced symbol error rate. Fig.~\ref{fig:ae_fig_example} shows the example loss distribution for samples with errors less than and greater than 20\% threshold and examples of out-of-distribution and within distribution hand pose samples. The two distributions have drastically different loss values, which can be used to infer the quality of the poses that can be displayed to the user. Additionally, we also retain pose samples that have fewer errors, which we correct.
It should be noted that the excluded hand pose samples are exceptions and does not represent the average pose received by \system. 
\begin{figure}[h]
	\centering
		\includegraphics[width=0.51\textwidth]{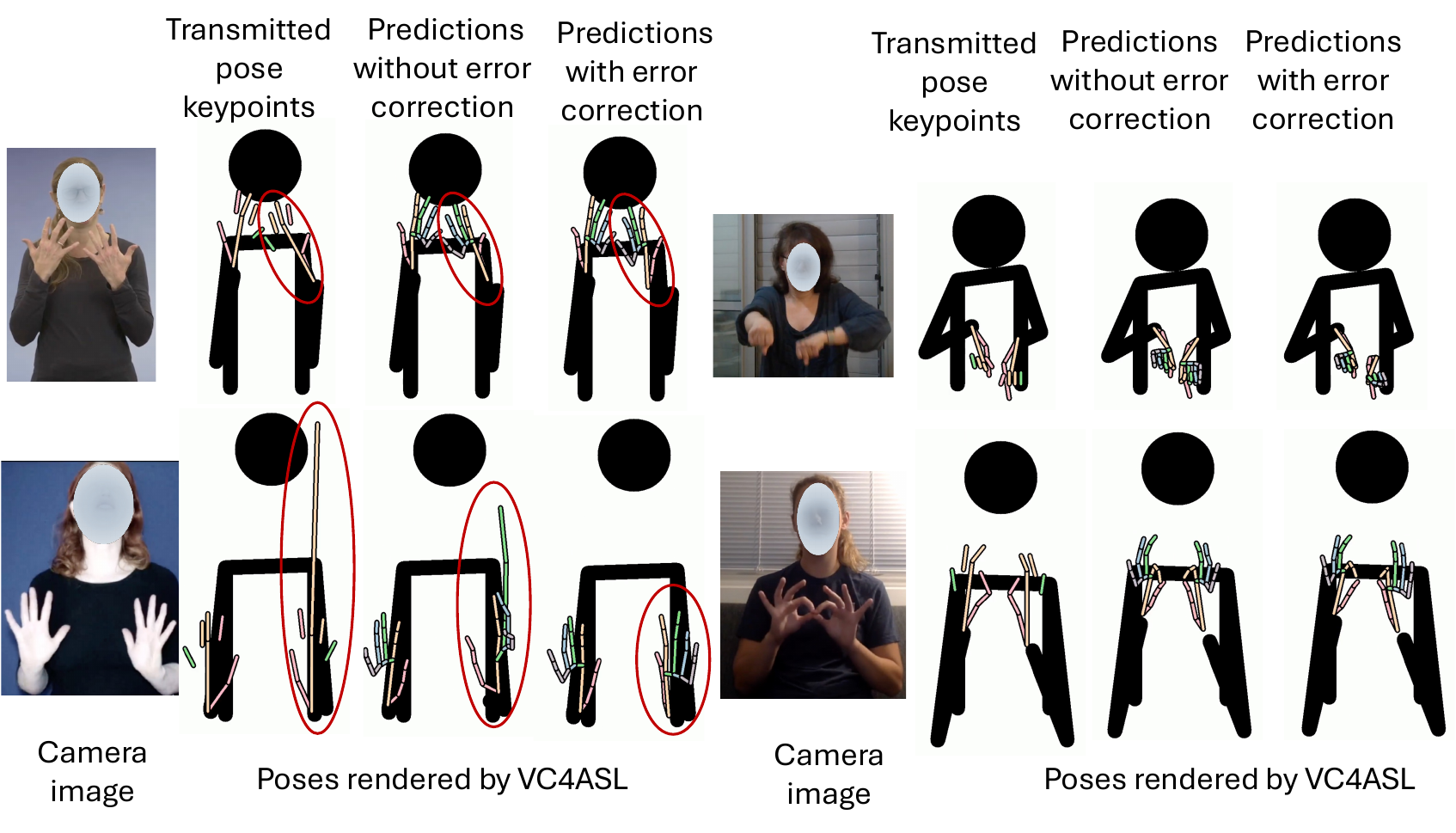}
		\label{fig:loss_distribution}
\vspace{-10pt}
	\caption{The poses rendered by \system from the transmitted keypoints for different ASL words. The errors in transmitted keypoints are highlighted with red ellipses.}
    \label{fig:rendered_poses}
		\vspace{-10pt}
\end{figure}
\subsubsection{Error Correction and Rendered Pose}In line with our low-latency design objectives, we retrain the excluded keypoint prediction model by incorporating expected noise characteristics during training. To tolerate up to $20\%$ symbol errors, at most two joint-level errors must be corrected per frame. Given that hand keypoints constitute approximately $73.33\%$ of all transmitted symbols, they are more likely to be affected by transmission errors. Moreover, as previously discussed, handshapes are critical to ASL communication, and errors in hand keypoints can have a more detrimental effect than errors in body keypoints. Accordingly, \system focuses exclusively on correcting hand keypoint errors. We model individual hand keypoint errors as samples from a Gaussian distribution with mean $\mu_e$ and standard deviation $\sigma_e$, and inject such noise into a subset of training samples within each batch. The parameters $\mu_e$ and $\sigma_e$ are empirically estimated using a standard reference pose by computing the discrepancy between ground-truth and received keypoints under varying network conditions. We use the Euclidean distance \cite{wikipediaEuclideanDistance} to quantify this discrepancy.

Finally, we use the error-corrected hand pose keypoints, along with the body pose keypoints, to render a fine-grained skeleton pose for the user. While the rendered skeleton is not equivalent to, nor a replacement for, the original camera images, it is designed to preserve key ASL-relevant features, such as handshape and body positioning. Fig.~\ref{fig:rendered_poses} presents examples of the final skeleton poses rendered for ASL users. For visual comparison, we include three versions: poses rendered without the excluded keypoints, those reconstructed using the excluded keypoint prediction model without error correction, and those reconstructed using the model trained with error correction. First, our rendering effectively captures critical ASL characteristics, such as relative head position, handshape, and finger distinction, using colored joints, yielding visual representations that closely resemble the camera-based images shown on the right. Second, compared to the model trained without error correction, the error-aware model demonstrates increased robustness to keypoint perturbations, particularly evident in columns two through four of both rows. Lastly, incorporating error correction during training reduces inference latency while improving prediction accuracy, especially for poses with minimal to no keypoint errors.

\section{Implementation and Evaluation}
\subsection{Implementation}
We implement the end-to-end prototype of \system on Windows laptops equipped with an Intel Core i7 processor and 16 GB of RAM on both the sender and receiver sides. The entire codebase is developed in Python. On the sender side, two separate processes are employed: one for the Pose Processor and another for the Pose-to-Audio Encoder. Python’s built-in Manager object \cite{pythonMultiprocessingProcessbased} is used to synchronize the transfer of quantized pose symbols between the Pose Processor and the Pose-to-Audio Encoder. The Pose Processor operates in a multi-threaded fashion, with a dedicated thread for inputting frames into the virtual camera interfaced with the video conferencing application (VCA) and three threads for Mediapipe-based pose estimation and subsequent processing. We utilize the Pose Landmarker (Heavy) and Hand Landmarker (Full) model bundles from Mediapipe \cite{mediapipe} for pose and hand keypoint estimation. Mediapipe contributes significantly to \system’s overall latency; approximately $70$–$100$ ms are spent on pose estimation. A similar setup is used on the receiver side, with two processes: one for the Pose-to-Audio Decoder and another for the Pose Renderer. 

\subsection{Evaluation and Numerical Results}

We evaluate {\system} by assessing both its individual components and conducting an end-to-end evaluation of the complete pipeline. Our evaluation aims to answer the following research questions:

\begin{itemize}
\item \textbf{Modulation Robustness:} How robust is the system’s modulation scheme under varying network conditions encountered during video conferencing?

\item \textbf{Effectiveness of Pose Rendering:} How do symbol errors introduced during transmission affect the accuracy of the reconstructed pose keypoints?

\item \textbf{Representation of ASL Components:} To what extent can the rendered poses accurately represent critical components of ASL, such as hand shapes and arm movements?

\item \textbf{Effectiveness for ASL Communication:} Can the poses rendered by {\system} support effective ASL communication for ASL speakers?
\end{itemize}

We address these questions through a series of evaluations, including symbol error rate measurements under controlled network conditions, pose keypoint distortion analysis, recognition experiments on ASL handshapes and words, and a user study involving both ASL and non-ASL participants. Together, these evaluations provide a comprehensive assessment of {\system}’s ability to support reliable ASL communication over video conferencing applications.

\subsubsection{Experimental Setup}

Unless otherwise stated, we follow the experimental protocol described below for all evaluations. We connect the sender and receiver through the selected video conferencing applications (VCAs), Zoom and Google Meet, using a dedicated 1 Gbps Ethernet link between the two hosts—a setup commonly used in prior measurement studies \cite{measurement_IMC,measurement_TON}.
Network conditions are varied separately for the sender and receiver sides using NetLimiter \cite{netlimiterNetLimiter}, as detailed in Section~\ref{sec:codec-impact}. In addition to the ideal no-limit scenario, we also simulate extreme network conditions by capping both sender and receiver speeds to 200 Kbps and 400 Kbps, where video communication typically fails. We limit the upload speeds on the sender side and the download speeds on the receiver side. On the sender side, we simulate ASL video conferencing by replaying predefined ASL signer videos, following an approach similar to \cite{measurement_IMC}. Depending on the specific evaluation, we vary the video input; further details are provided in the respective sections. The system captures frames from the host machine, processes them through the sender-side pipeline, and outputs the corresponding images and modulated audio to the virtual camera and virtual microphone interfaces. This processing occurs in real time. On the receiver side, we capture both the video and audio streams, which are subsequently processed offline to evaluate system performance. Additionally, since Zoom defaults to filtering our modulation, we use the setting for live performance \cite{zoomConfiguringAudio}.
\begin{figure}[h]
	\vspace{-15pt}
	\centering
    \captionsetup[subfloat]{labelfont=scriptsize,textfont=scriptsize}
	\hspace*{-10pt}
	\subfloat[]{
		\includegraphics[width=0.25\textwidth]{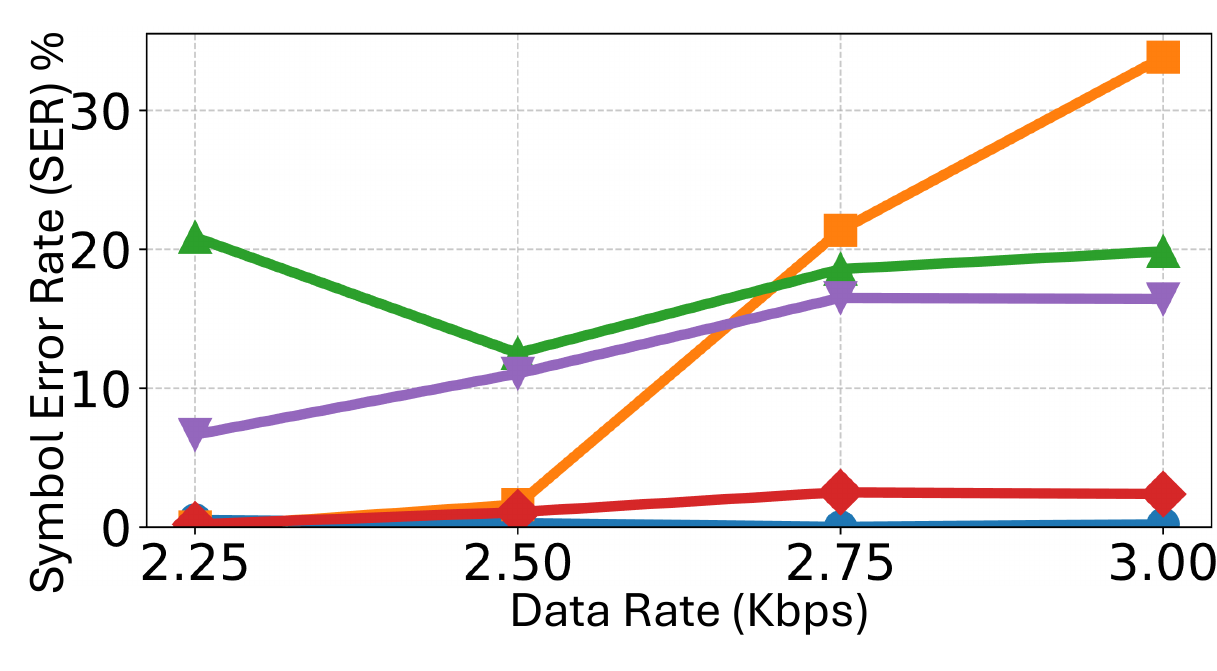}
		\label{fig:zoom_ser_modulation}

	} 
    		\hspace*{-8pt}
	\subfloat[]{
		\includegraphics[width=0.252\textwidth]{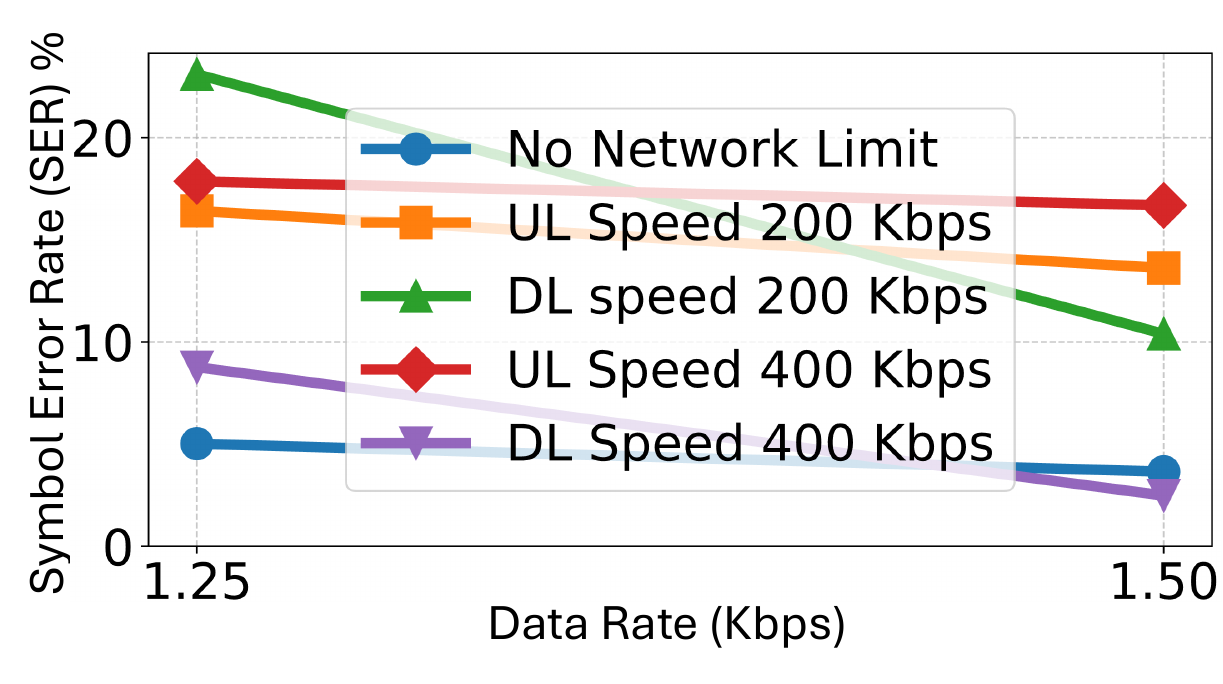}
		\label{fig:meet_ser_modulation}
	} 
	\vspace{-5pt}
	\caption{Evaluating {\system}'s modulation scheme under different network configurations and data rates.(a) SER for Zoom under different network and data rate settings and (b) SER for Meet under different network and data rate settings}
    \label{fig:modulation-ser}
		\vspace{-10pt}
\end{figure}

\subsubsection{Modulation Robustness}

We evaluate the Symbol Error Rate (SER) of \system’s modulation scheme by transmitting a fixed, uniformly random sequence of symbols via the audio channel and measuring the percentage of incorrectly received symbols. Unlike prior evaluations where we controlled codec parameters directly, here we assess the impact of video conferencing applications (VCAs) under codec settings that dynamically adapt to network conditions. To maintain consistency with the earlier setup, we feed ASL video frames into the virtual camera interface of the VCA during transmission. Experiments are conducted across varying data rates and network speed limits on both the sender and receiver sides. For each configuration, we transmit data for five minutes and collect five independent samples to compute the average SER. Fig.~\ref{fig:modulation-ser} reports SER across configurations for Zoom and Google Meet. The results show that the modulation scheme exhibits different behaviors depending on the VCA and network condition, with significant performance differences between Zoom and Meet.

Meet runs in a web browser and applies built-in noise filtering, which we could not disable. This filtering contributes to higher SERs at data rates exceeding 1.5 Kbps—the upper limit of our modulation scheme. In contrast, Zoom’s SER increases consistently with both higher data rates and lower network speeds, reaching up to 33\%, when the sender upload is capped at 200 Kbps. At this bandwidth, Zoom aggressively adapts its codec settings, particularly on the sender side. Meet exhibits a similar trend but peaks at a lower SER of 23.1\%, observed when the receiver’s download speed is limited.
These differences stem from the distinct codec adaptation strategies employed by each VCA. Overall, \system tolerates extreme network conditions by reducing data rates, highlighting a tradeoff between reliability (SER) and throughput under constrained bandwidth.

\textbf{Impact on Received Pose Keypoints:} To assess the relationship between symbol error rate (SER) and the quality of received pose keypoints, we replicate the SER experiments under similar network and data rate configurations. Instead of transmitting a random symbol sequence, we repeatedly send a static pose depicting a user with raised arms and all fingers visible. This pose is processed through the full system pipeline and received on the other end, where the output is compared against the ground-truth pose using Euclidean distance in pixel space as the joint error metric. We report average joint errors separately for body and hand keypoints.
\begin{figure}[h]
	\centering
    \captionsetup[subfloat]{labelfont=scriptsize,textfont=scriptsize}
	\hspace*{-10pt}
	\subfloat[Pose errors in Zoom]{
		\includegraphics[width=0.52\textwidth]{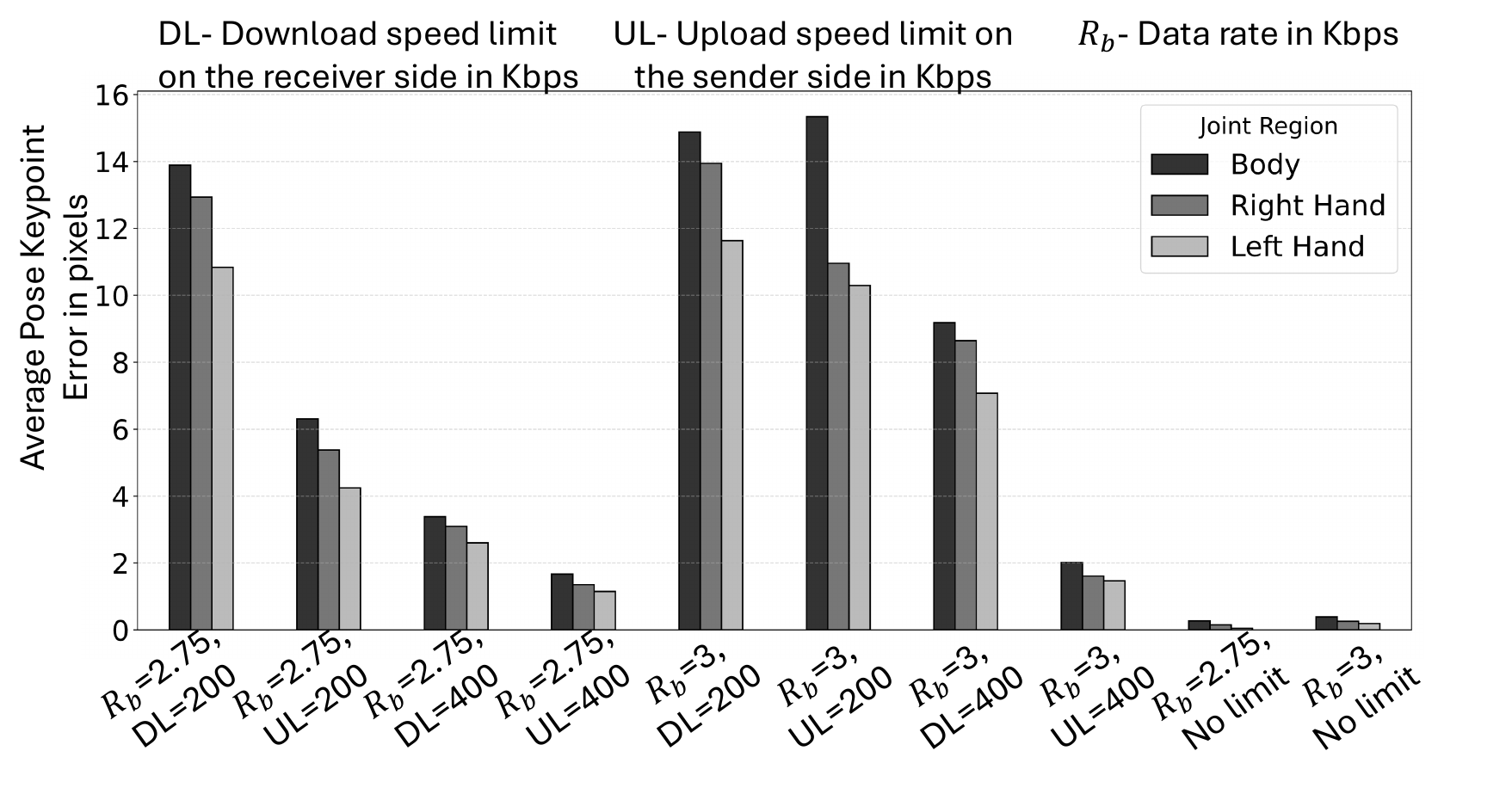}
		\label{fig:zoom_pose_error}
		\hspace*{-5pt}
	} 
    \\
	\subfloat[Pose errors in Meet]{
		\includegraphics[width=0.52\textwidth]{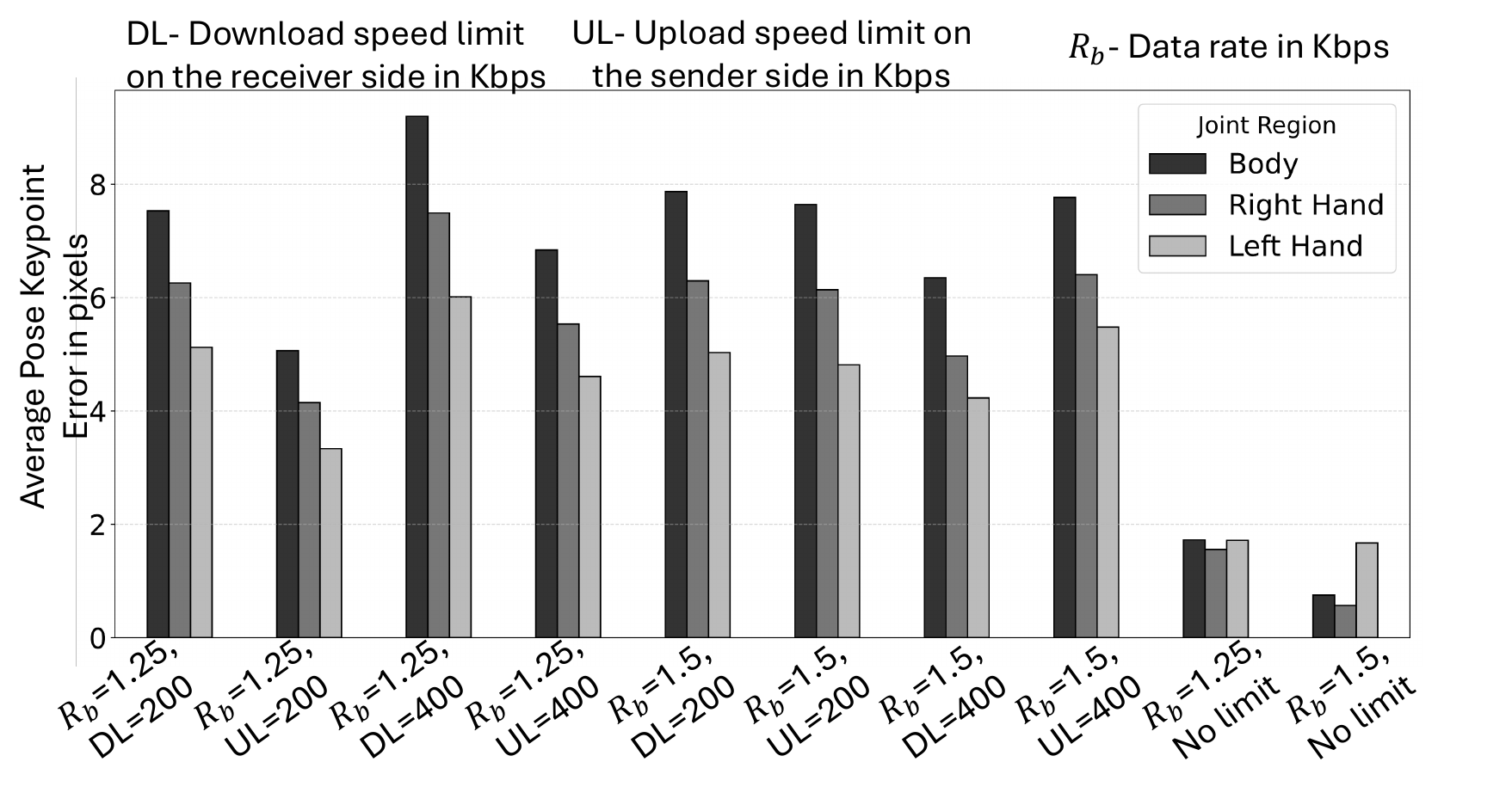}
		\label{fig:meet_pose_error}
	} 
	\vspace{-5pt}
	\caption{Pose keypoints errors using \system under varying network conditions and data rates.}
		\vspace{-10pt}
        \label{fig:pose-keypoint-errors}
\end{figure}

Fig.~\ref{fig:pose-keypoint-errors} presents the results across different settings. Consistently, body keypoint errors exceed those of the hands, which is expected given the larger spatial displacement between body joints. Error trends closely mirror those observed in SER, with higher data rates and constrained network bandwidth leading to greater degradation. Notably, even with a peak SER of 33\% (Zoom), the corresponding average body keypoint error remains within 15.24 pixels—suggesting potential for effective post hoc error correction. Under ideal conditions (bandwidth $>$ 200 Kbps), Zoom outperforms Meet, even at elevated data rates. Across all conditions, the joints with the highest errors are the nose (body) and wrists (hands), consistently across both platforms. We hypothesize that these joints, occupying fixed positions in {\system}'s frame, are particularly susceptible to codec-induced distortion.

\begin{table}[h]
\centering
\caption{Performance of Pose Error Detection}
\begin{tabular}{lcc}
\toprule
\multirow{2}{*}{\textbf{Metric}} & \multicolumn{2}{c}{\textbf{Error Detection Method}} \\
\cmidrule(lr){2-3}
 & \textbf{PCA-based} & \textbf{VC4ASL's} \\
\midrule
Accuracy (\%)  & 84.11 & 86.60 \\
F1-Score (\%)  & 84.20 & 86.60 \\
\bottomrule
\end{tabular}
\label{tab:error-detection-performance}
\end{table}

\subsubsection{Effectiveness of Pose Rendering}

\system’s pose rendering pipeline includes two components: an error detection module and a pose keypoint prediction model with error correction. To evaluate their effectiveness, we train and test these components using the How2Sign dataset \cite{how2sign}, as described in Section~\ref{sec:point-detection}.

\textbf{Error Detection Performance:} Since no labeled ground truth exists for training or evaluating the autoencoder-based error detection model, we synthetically generate labels by injecting noise into the pose keypoints. Specifically, we introduce random perturbations exceeding 20\% of the joint coordinates, marking these as erroneous samples, and treat unperturbed samples as non-erroneous. As a baseline, we compare the autoencoder with a Principal Component Analysis (PCA)-based method. PCA has been widely applied in denoising tasks through low-rank reconstruction \cite{babu2012pca}. Following this approach, we fit a PCA model to the training set and compute the reconstruction error on test samples. We then apply a threshold—analogous to the autoencoder setting—to classify samples as erroneous or not, based on their reconstruction loss.

We evaluate performance using Accuracy and F1-Score. Table~\ref{tab:error-detection-performance} summarizes the results, showing that the proposed autoencoder-based method outperforms the PCA-based baseline on both metrics. Given the anomaly detection nature of the task, a higher F1-Score reflects a better balance between precision and recall. Notably, the PCA-based method still achieves competitive performance, demonstrating its viability as a lightweight baseline.

\begin{table*}[ht]
\centering
\caption{Comparison of pose keypoint prediction performance. \\
}
\label{tab:pose_prediction}
\resizebox{\textwidth}{!}{%
\begin{tabular}{@{}lcccc@{}}
\toprule
\textbf{Error Metric} & \textbf{Linear Interpolation} & \textbf{Linear Regression} & \textbf{VC4ASL w/o Error Correction} & \textbf{VC4ASL w/ Error Correction} \\
\midrule
Mean Absolute Error (MAE)         & 0.9789 & 0.2249 & 0.2900 & 0.2970 \\
Mean Squared Error (MSE)          & 2.5410 & 0.2004 & 0.1710 & 0.1888 \\
Coefficient of Determination ($R^2$)  & 0.9720 & 0.9960 & 0.9970 & 0.9960 \\
\bottomrule
\end{tabular}%
}
\end{table*}

\textbf{Excluded Keypoint Prediction and Error Correction:}
We evaluate two baselines: linear interpolation of pose keypoints and a linear regression model trained on the same How2Sign data as \system. Both are tested on the same held-out set. Performance is assessed using standard regression metrics, Mean Squared Error (MSE), Mean Absolute Error (MAE), and $R^2$ score, comparing predicted and ground-truth keypoints. For \system, we compare models trained with and without error correction, where noise is introduced during training by randomly perturbing pose keypoints.

Table~\ref{tab:pose_prediction} reports the results for \system and the baselines. Compared to linear interpolation, which often loses structural consistency, \system yields lower MSE and MAE and a higher $R^2$, indicating better fidelity. Interestingly, the model trained with error correction shows slightly higher errors than the one trained without, suggesting that incorporating noise during training has a minimal adverse effect. The linear regression baseline achieves performance close to \system and may serve as a viable lightweight alternative in resource-constrained settings.

\begin{figure*}[t]
	\vspace{-5pt}
	\centering
    \captionsetup[subfloat]{labelfont=scriptsize,textfont=scriptsize}
	\hspace*{-10pt}
	\subfloat[Impact of noise on excluded keypoint predictions ]{
		\includegraphics[width=0.38\textwidth]{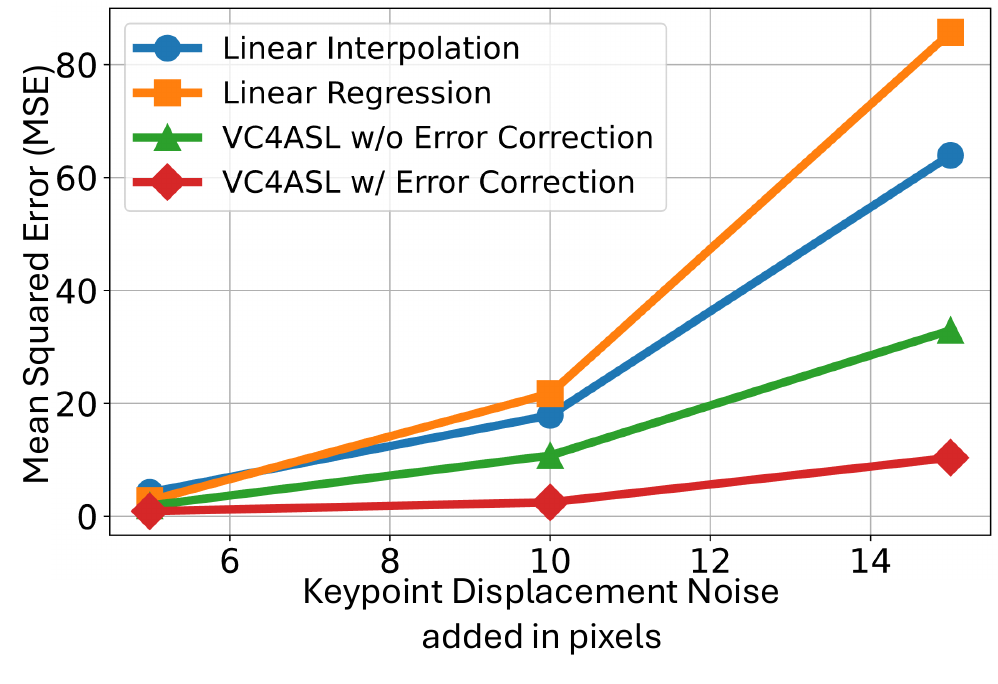}
		\label{fig:vc4asl_noise_impact}

	} 
    		\hspace*{-10pt}
	\subfloat[Percentage of erroneous frames under different configurations]{
		\includegraphics[width=0.52\textwidth]{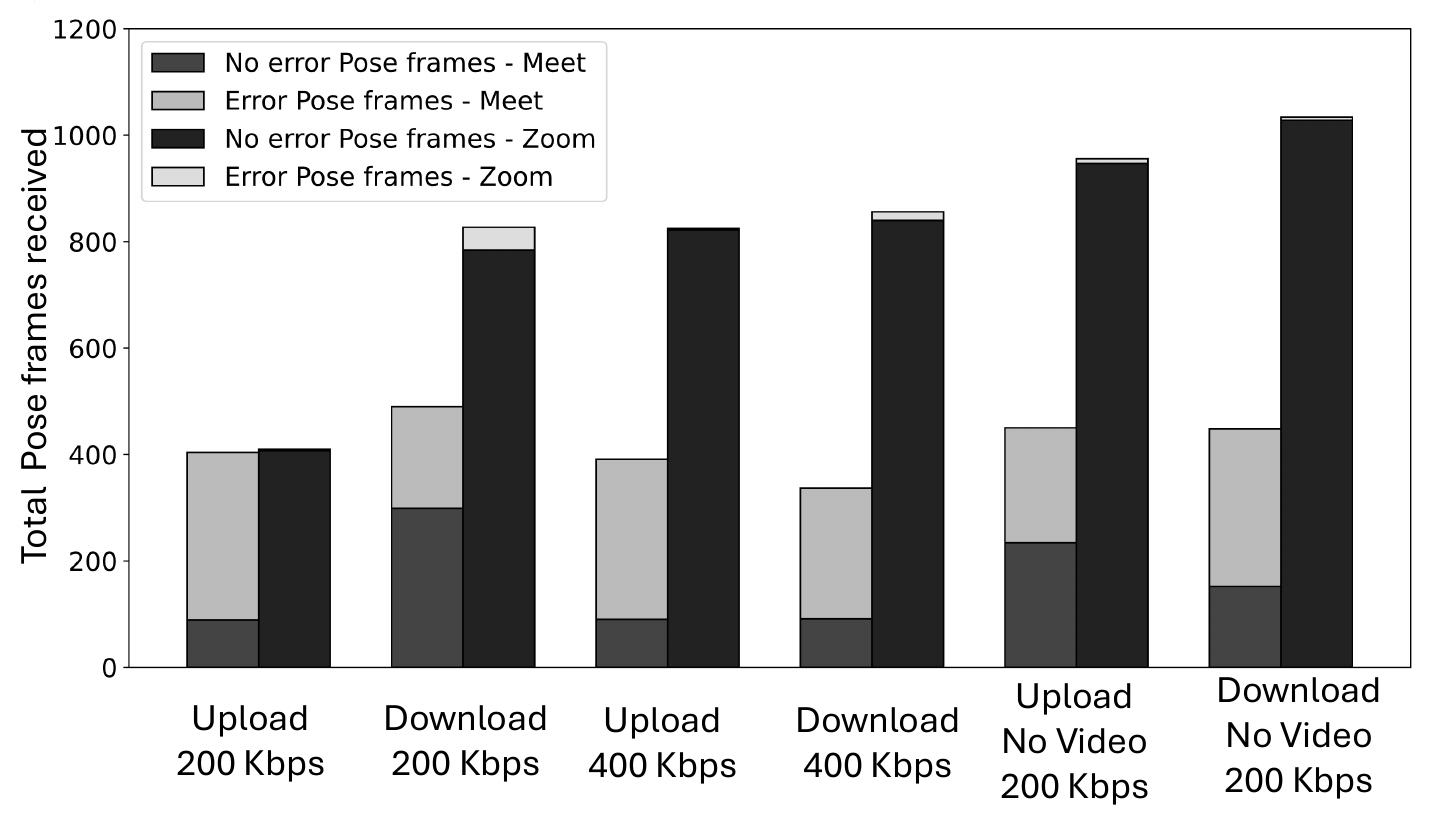}
		\label{fig:errors_under_network}
	} 
	\vspace{-5pt}
	\caption{Evaluating {\system}'s error correction model and the impact of network conditions on the percentage of errors in received pose frames.}
    \label{fig:error-correction}
		\vspace{-10pt}
\end{figure*}

To analyze robustness to transmission-induced errors, we introduce synthetic noise in the test set by displacing the pixel positions of one or two joints, and evaluate predictions on the excluded keypoints. Fig.~\ref{fig:vc4asl_noise_impact} shows the results across different noise levels. Both \system models exhibit strong error resilience relative to the baselines, with the error-corrected model maintaining performance even under severe perturbations. However, we observe that \system may introduce minor distortions to uncorrupted input keypoints, a side effect of its joint reconstruction objective. In contrast, baseline linear models retain input keypoints exactly but fail to generalize under noisy conditions. This highlights the potential of hybrid approaches that combine the robustness of non-linear models with the stability of linear ones.
\subsubsection{Representation of ASL Components}

To evaluate \system’s ability to accurately convey ASL-specific attributes, such as handshape, relative body positioning, and arm movement, we conduct two end-to-end machine recognition tasks: ASL alphabet recognition \cite{asl_alphabet} and isolated sign recognition \cite{WLASL}. The alphabet recognition task focuses on handshape fidelity, as each ASL letter corresponds to a distinct static hand configuration. Success in this task reflects the precision of the rendered hand keypoints. The isolated sign recognition task, on the other hand, evaluates the system’s ability to preserve dynamic temporal features, such as joint displacements and inter-limb coordination, which are critical for recognizing motion-based ASL signs.

For both tasks, we collect evaluation data under previously defined network conditions, using fixed data rates of $R_b = 3$ Kbps for Zoom and $R_b = 1.5$ Kbps for Meet, based on earlier findings. Due to Meet’s lower data rate, we perform the isolated sign recognition task exclusively on Zoom, where the available bandwidth supports more accurate temporal pose rendering.

\begin{table*}[htbp]
\centering
\small
\renewcommand{\arraystretch}{1.3}
\setlength{\tabcolsep}{4pt}
\caption{Performance of ASL alphabet detection under different network conditions for Meet and Zoom. UL- Upload, DL- Download.}
\begin{tabular}{lcccccccc}
\toprule
\textbf{Metric} & \multicolumn{2}{c}{\textbf{UL 200 Kbps}} & \multicolumn{2}{c}{\textbf{DL 200 Kbps}} & \multicolumn{2}{c}{\textbf{UL 400 Kbps}} & \multicolumn{2}{c}{\textbf{DL 400 Kbps}} \\
 & Meet & Zoom & Meet & Zoom & Meet & Zoom & Meet & Zoom \\
\midrule
Accuracy Without Error Detection (\%)  & 17 & 94 & 46 & 94 & 13 & 97 & 19 & 96 \\
Accuracy With Error Detection (\%)     & 96 & 94 & 95 & 96 & 96 & 97 & 97 & 97 \\
Percentage Errors Detected           & 78 & 0 & 41 & 2 & 80 & 0 & 75 & 1 \\
\bottomrule
\end{tabular}

\label{tab:alphabet_detection}
\end{table*}

\textbf{Alphabet Recognition:}
We evaluate \system using a publicly available Kaggle dataset and pre-trained model \cite{kaggleRecognitionMediapipe}, following the original train-test split. Two letters, \texttt{J} and \texttt{Z}, are excluded due to their motion-based representation. On the sender side, we play static handshape frames and transmit all hand keypoints through the \system pipeline, without applying error correction, to isolate the impact of the modulation scheme. Accuracy is used as the evaluation metric. As we cannot know the exact pose frames received,  we manually label the received pose frames to generate the ground truth.

Table~\ref{tab:alphabet_detection} reports results for Meet and Zoom, with and without error detection. When erroneous frames are excluded using the error detection model, accuracies reach 95.8\% (Meet) and 96.2\% (Zoom). Without error detection, Meet’s accuracy drops by over 70\%, while Zoom exhibits a smaller decline. This aligns with the higher rate of erroneous frames observed in Meet due to its aggressive noise filtering.

To further quantify these differences, we compare the number of received frames and the proportion dropped as erroneous. Fig.~\ref{fig:errors_under_network} shows that Zoom generally receives more frames due to its higher data rate. However, under a 200 Kbps upload cap, Zoom frequently disconnects, halving the number of frames received. Disabling video under poor network conditions mitigates this issue, increasing frame count from 413 to 1040. Meet, by contrast, does not exhibit this behavior due to its fixed transmission strategy. However, using \system over Meet can lead to up to 78\% erroneous frames under low bandwidth, due to the browser's built-in noise filtering. Overall, Zoom provides more reliable support for \system, and is exclusively considered for subsequent evaluations.

\textbf{ASL Word Recognition:} We select a subset of $13$ words from the WLASL dataset \cite{WLASL} and retrain the original model using only this subset. We use 80\% of the data for training and the remaining 20\% for testing. Each selected word is associated with distinct handshapes and movements, and the reduced vocabulary enables controlled experimentation across different network and system conditions. To evaluate performance, we generate a video from the test set words and pass it through {\system}’s pipeline. On the receiver side, the output is manually segmented and labeled to create ground truth pose sequences.

We compare two models: one trained with error correction and one without. Table~\ref{tab:vc4asl_accuracy} summarizes the recognition results under these configurations. Both models demonstrate comparable performance across the selected ASL words, indicating that the rendered pose keypoints after the excluded keypoint prediction preserve the temporal continuity. However, the model trained without error correction shows a noticeable drop in performance. In contrast, the model incorporating error correction better preserves structural cues,  such as hand shape, which are critical for ASL recognition, thereby achieving superior accuracy.

\begin{table*}[htbp]
\centering
\footnotesize
\renewcommand{\arraystretch}{1.7}
\setlength{\tabcolsep}{6pt}
\caption{ASL recognition accuracy of Vc4ASL received pose keypoints under various network conditions.}
\begin{tabular}{lcccc}
\toprule
\multicolumn{5}{c}{\textbf{Accuracy (\%)}} \\
\midrule
\textbf{Model} & \textbf{No Network Limit} & \textbf{Download 200 Kbps} & \textbf{Download 400 Kbps} & \textbf{Upload 400 Kbps} \\
\midrule
Vc4ASL without Error Correction & 90.52 & 81.23 & 90.12 & 90.51 \\
Vc4ASL with Error Correction    & 100.00 & 81.82 & 90.92 & 100.00 \\
\bottomrule
\end{tabular}
\label{tab:vc4asl_accuracy}
\end{table*}
\subsubsection{Effectiveness for ASL Communication}

We assess the usability of \system with American Sign Language (ASL) speakers through an IRB-approved user study. Participants were recruited from the ASL community, and the study description stated that the goal was to examine the impact of ASL communication on video conferencing applications (VCAs), with participation being entirely voluntary. The study was conducted using a Google Form that included six videos; all except the first were followed by a prompt asking participants to transcribe the ASL content in gloss form. The first video served as an introduction, presenting both camera and pose frames side-by-side, along with subtitles. This introductory video helped participants build an understanding of how pose-based representations correspond to conventional ASL video, facilitating comprehension in the following videos.

The next four videos consisted solely of pose frame sequences generated by \system under different network conditions using Zoom. Each video was constructed from a subset of 14 commonly used ASL sentences, arranged in different orders to mitigate familiarity bias. The final video was generated directly from MediaPipe pose keypoints without passing through the video conferencing application. This served as a control condition to evaluate comprehension under ideal pose rendering without transmission-induced distortions.

\begin{table}[h]
\centering
\footnotesize
\renewcommand{\arraystretch}{1.4}
\setlength{\tabcolsep}{3pt}
\caption{Semantic Similarity Scores under different network conditions.}
\begin{tabular}{lccccc}
\toprule
\multirow{2}{*}{\textbf{User}} 
 & \multicolumn{2}{c}{\textbf{200 Kbps}} 
 & \multicolumn{2}{c}{\textbf{400 Kbps}} 
 & \multirow{2}{*}{\textbf{No VCA}} \\
\cmidrule(lr){2-3} \cmidrule(lr){4-5}
 & Download & Upload & Download & Upload & \\
\midrule
User 1 & 0.43 & 0.68 & 0.60 & 0.56 & 0.85 \\
User 2 & 0.71 & 0.79 & NA   & 0.92 & 0.88 \\
User 3 & 0.58 & 0.79 & 0.70 & 0.82 & 0.82 \\
User 4 & 0.72 & 0.78 & 0.76 & 0.77 & 0.86 \\
User 5 & 0.46 & 0.68 & 0.45 & 0.71 & 0.83 \\
User 6 & 0.36 & 0.66 & 0.63 & 0.55 & 0.78 \\
User 7 & 0.55 & 0.74 & 0.67 & 0.78 & 0.79 \\
User 8 & 0.39 & 0.74 & 0.67 & 0.79 & 0.79 \\
\bottomrule
\end{tabular}
\label{tab:semantic-similarity}
\end{table}
\textbf{ASL Comprehension from Pose Videos:}
To estimate user comprehension, we use the semantic similarity score \cite{wikipediaCosineSimilarity}, defined as the cosine similarity \cite{wikipediaCosineSimilarity} between the embedding vectors of the ground truth and the user’s response. Semantic similarity captures the conceptual relatedness between two texts and avoids the limitations of lexical similarity, which relies on direct word-for-word comparison. We use Sentence Transformers \cite{sbertSentenceTransformersDocumentation} to generate sentence-level embeddings for each user’s full response to a video and compare them against the corresponding ground truth embedding. This simulates a conversational context in which users are expected to interpret longer segments before responding.

Table~\ref{tab:semantic-similarity} presents the semantic similarity scores across eight users under different network conditions. In the absence of a video conferencing application (VCA), shown in the last column, users had little difficulty understanding the ASL videos, with an average similarity score of 0.825. However, substantial variation is observed across users in the other conditions. For instance, User 1 consistently achieved lower scores in all VCA-enabled settings, while User 2 consistently achieved higher scores across videos. Notably, User 2 was unable to comprehend the video under the 400 Kbps download limit and remarked, “Cannot see a lot of glitches in this video.” This video included several erroneous pose frames, which we intentionally retained to evaluate their impact on comprehension, an issue that other users appeared less affected by. Additionally, we observe that most users obtained lower semantic scores for the video with a 200 Kbps download limit, with an average score of 0.59. We attribute this to the fact that this was the first video presented in the sequence, during which users were still adapting to the pose-based ASL representation. Comprehension generally improved in subsequent videos.

\begin{table*}[htb]
\centering
\small  
\renewcommand{\arraystretch}{1.2}
\setlength{\tabcolsep}{6pt} 
\caption{Comparison of ASL gloss with user responses}
\begin{tabular}{>{\raggedright\arraybackslash}p{3.5cm} >{\raggedright\arraybackslash}p{3.2cm} >{\raggedright\arraybackslash}p{3.2cm} >{\raggedright\arraybackslash}p{3.2cm}}
\toprule
\textbf{Original ASL Gloss} & \textbf{User Response 1} & \textbf{User Response 2} & \textbf{User Response 3} \\
\midrule
Good job keep going         & Good job, next.          & Good work, go ahead.     & Good work ahead!         \\
Do you mind write           & I like to write.         & Mind writing?            & You write?               \\
You name what you           & You name what?           & You name what you?       & What’s your name?        \\
You work where you          & You work where you?      & Where do you work at?    & Work—where?              \\
Me hungry restaurant you want goto you & I'm hungry. Do you want to go to a restaurant? & I'm hungry. Do you want to go to a restaurant? & I’m hungry for a restaurant. Do you want to go? \\
\bottomrule
\end{tabular}

\label{table:asl-gloss-table}
\end{table*}

\textbf{Understanding the Variance:}
The sentence set included a range of words requiring diverse handshapes and finger movements. Notably, three finger-spelled words, \texttt{Bob}, \texttt{ASL}, and \texttt{Cup}, were comprehended by users with average accuracies of 75\%, 95.83\%, and 71.85\%, respectively, across all settings. These results suggest that \system’s predicted pose keypoints can effectively convey handshape, which is critical for accurate fingerspelling. One common error was misinterpreting \texttt{Bob} as \texttt{Fob}, a plausible confusion given the visual similarity between the two signs \cite{wikipediaAmericanManual}.

Additionally, we observed that Users 5 and 6 submitted some of their responses in English rather than ASL gloss, which contributed to lower semantic similarity scores. ASL glosses annotate individual signs and typically differ in structure from English sentences, making direct comparisons less straightforward. Table~\ref{table:asl-gloss-table} presents a subset of sentences, their original glosses, and user responses across different conditions. While variations exist between user responses, most convey meanings consistent with the original glosses.

Overall, the findings indicate that while there remains room for improvement, \system in its current form demonstrates the potential to facilitate ASL communication under low-bandwidth network conditions.

\section{Related Work}
We categorize related work into three areas that align with the key components of \system.
\subsection{Encoding Data in Audio}
Prior research has explored a variety of methods for encoding data in audio, particularly from an information-hiding perspective. Temporal techniques such as Least Significant Bit (LSB) substitution embed information over time by altering the least significant bits \cite{tian2008covert}. In contrast, modulation-based methods, such as phase, frequency, and amplitude modulation, provide greater resilience to codec-induced distortion, though each comes with its own trade-offs in robustness and capacity  \cite{nutzinger2011novel, takahashi2007assessment, deng2008novel}. Some approaches modify audio codecs themselves, but these tend to be less practical due to compatibility issues. Additionally, voice-specific encoding schemes, such as those based on Packet Loss Concealment (PLC) algorithms, can support transmission rates up to 400 bits/sec \cite{geiser2008steganographic}. Unlike steganographic methods that aim to conceal information, \system is designed to achieve higher data rates while maintaining robustness against codec and network impairments. To this end, \system leverages Chirp Spread Spectrum (CSS) modulation to achieve reliable and high-throughput transmission.

\subsection{Low Data Rate Video Communication}

Compression remains a standard approach for enabling low-data-rate video communication. Prior work has extensively investigated low-bitrate compression methods \cite{marpe1999very, rijkse2002h}, several of which have been adopted as video coding standards \cite{zhang1995very}. More recently, the emergence of generative models has opened new avenues for transmitting minimal information and reconstructing full video frames on the receiver side. For example, \cite{wang2021one} proposes synthesizing a person’s talking head using limited inputs. Similarly, \cite{guo2021ad} and \cite{li2024talkinggaussian} employ audio-driven neural radiance fields and Gaussian splatting techniques for talking head generation. While person image synthesis from pose inputs has also been explored \cite{ma2017pose}, the quality of synthesized hand regions remains a significant limitation. In the context of sign language, researchers have investigated both gloss-to-pose and pose-to-sign generation pipelines \cite{sharma2024sign}. These emerging image generation techniques could augment our approach by reconstructing ASL signer visuals at the receiver end, thereby addressing the intelligibility challenges inherent in pose-only ASL video communication.

\subsection{ASL Communication in Video Conferencing Applications}

ASL users encounter a range of accessibility barriers when engaging with mainstream video conferencing applications. Prior work has documented several of these challenges, including difficulties in gaining others’ attention, interpreter delays, and video quality limitations that hinder accurate sign interpretation \cite{asl_accessibility_1}. In \cite{asl_accessibility_4}, the authors introduced Erato, a tool designed to provide data-driven socio-technical support for Deaf and Hard of Hearing presenters. Additionally, \cite{asl_accessibility_2} describes a co-design approach to developing video conferencing tools tailored for collaborative environments involving users with mixed hearing status. Collectively, these efforts highlight the accessibility gaps present in current video conferencing platforms and propose solutions aimed at enhancing inclusivity. Complementing this body of work, our approach addresses a core technological limitation faced by ASL users in low-bandwidth settings, offering a solution that operates robustly under constrained network conditions.
\section{Discussion and Limitations}

\begin{itemize}
\item Owing to bandwidth limitations, \system currently transmits only hand and body pose keypoints. However, facial expressions constitute an integral component of ASL grammar, and their absence may impact the fluency and completeness of signed communication. This constraint primarily arises from the need to maintain robustness against codec-induced distortions and fluctuating network conditions. Future extensions of \system could incorporate facial keypoints using existing face mesh estimation methods, provided the transmission channel can accommodate the increased data requirements.

\item The present implementation of \system does not incorporate a feedback mechanism between the sender and the receiver. Introducing such a mechanism could facilitate adaptive control over transmission parameters, such as symbol duration or modulation rate, thereby reducing error rates and improving overall communication reliability. In particular, the proposed error detection strategy can be leveraged to trigger adaptive behavior when the error rate exceeds a predefined threshold.

\item Although the rendered poses provide intelligible visual information for ASL interpretation, users may require a brief period of familiarization to become proficient in this communication modality. To further enhance usability, future work may explore image synthesis techniques to reconstruct photorealistic user representations on the receiver end. This would enable dynamic transitions between actual video frames and synthetic pose renderings during network degradation, improving visual continuity.

\item While \system is tailored for ASL communication, the underlying modulation framework is general and can be extended to support other forms of visual or symbolic communication through audio channels. This opens up broader applications in accessibility-focused technologies and bandwidth-constrained communication scenarios.
\end{itemize}

\section{Conclusion}

In this work, we introduced \system, a system that enables ASL communication over the audio channels of existing video conferencing applications. By encoding human pose data into audio signals and incorporating receive-side error detection and correction, \system facilitates communication when network conditions degrade video quality. Notably, the system integrates seamlessly with existing platforms without requiring any modifications. Through a series of component-level evaluations and user studies, we demonstrated that \system supports effective ASL communication, even in bandwidth-constrained environments.
\bibliographystyle{ieeetr} 
\bibliography{bibs}

\begin{IEEEbiographynophoto}{Panneer Selvam Santhalingam}
received his B.E. in Electrical and Electronics Engineering from SSN College of Engineering in Chennai, India, his M.S. in Information Security and Assurance from George Mason University, and his Ph.D. in Computer Science from George Mason University. He is currently an Assistant Professor in the Department of Computer and Information Science at Brooklyn College, City University of New York. His research interests include wireless and mobile computing, sensing, and next-generation wireless networks, with a focus on robustness and scalability in emerging sensing technologies.
\end{IEEEbiographynophoto}
\vspace{-30pt}
\begin{IEEEbiographynophoto}{Swann Thantsin}
 is a graduate student pursuing an M.S. in Computer Science at Brooklyn College, City University of New York. He works as a Software Research Assistant, helping students understand core computer science concepts (especially Java), and contributing to academic support roles.
\end{IEEEbiographynophoto}
\vspace{-30pt}
\begin{IEEEbiographynophoto}{Ahmad Kamari}
is a Ph.D. student in the Department of Computer Science at George Mason University. His research focuses on next-generation wireless networks, including millimeter-wave communications, RF sensing, and wireless networking protocols.
\end{IEEEbiographynophoto}
\vspace{-30pt}
\begin{IEEEbiographynophoto}{Parth Pathak}
received his Ph.D. in Computer Science from North Carolina State University in 2012. He was a postdoctoral researcher at the University of California, Davis, before joining George Mason University, where he is currently an Associate Professor in the Computer Science Department in the School of Computing. He is a recipient of the NSF CAREER Award (2021), CCI Fellowship (2023), IDIA P3 Fellowship (2023), and the Excellence in Postdoctoral Research Award at UC Davis (2015). His research interests are in wireless networking and mobile computing, with current work focusing on millimeter-wave wireless networks, machine learning for wireless networking and sensing, low-power high-speed IoT systems, and wireless sensing and imaging for autonomous robotic platforms.
\end{IEEEbiographynophoto}
\vspace{-30pt}
\begin{IEEEbiographynophoto}{Kenneth DeHaan}
 holds an Ed.D. in Higher Education Management (2020) from the University of Pittsburgh, an M.A. in Sign Language Education from Gallaudet University (2015), an M.S. in Professional Studies (Business, Service \& Tourism Management) from Rochester Institute of Technology (2011), a B.S. in Business Administration (2009), and an associate degree in Business Management (2007) from RIT. He is an Assistant Professor in American Sign Language at Gallaudet University and serves as Director of the Master's in Sign Language Education program, overseeing curriculum, faculty coordination, and student mentorship. His work is focused on improving ASL teacher preparation programs and advancing Sign Language Education, especially at the higher-education level.
\end{IEEEbiographynophoto}



\end{document}